\documentclass[aps,prc,superscriptaddress,twoside,twocolumn,nofootinbib,showpacs,showkeys]{revtex4-1}
\usepackage{amsmath,amssymb}
\usepackage{mathtools}
\usepackage{bbold}
\usepackage[usenames, dvipsnames]{xcolor}
\usepackage{braket}
\usepackage{graphicx,bm}
\usepackage{slashed}
\usepackage{tensor,siunitx}

\usepackage{changes}
\definechangesauthor[name={Yeunhwan}, color=blue]{YH}
\definechangesauthor[name={Yongseok}, color=red]{YO}
\definechangesauthor[name={Chang Hwan}, color=green]{CL}

\allowdisplaybreaks

\begin{document}

\title{\boldmath
Effective interactions of hyperons and mass-radius relation of neutron stars}

\author{Yeunhwan \surname{Lim} }
\email{ylim@tamu.edu}
\affiliation{Cyclotron Institute and Department of Physics and Astronomy, 
Texas A\&M University, College Station, Texas 77843, USA}

\author{Chang-Hwan \surname{Lee}}
\email{clee@pusan.ac.kr}
\affiliation{Department of Physics, Pusan National University, Busan 46241, Korea}

\author{Yongseok \surname{Oh}}
\email{yohphy@knu.ac.kr}
\affiliation{Department of Physics, Kyungpook National University, Daegu 41566, Korea}
\affiliation{Asia Pacific Center for Theoretical Physics, Pohang, Gyeongbuk 37673, Korea}

\begin{abstract}

We examine the role of hyperons in a neutron star based on the relativistic mean field approach.
For nuclear matter below 1.5 times the normal nuclear density we constrain the model parameters by
using the symmetric nuclear matter properties and theoretical investigations for neutron matter in the literature.
We then extend the model to higher densities by including hyperons and isoscalar vector mesons 
that contain strangeness degree of freedom.
We confirm that the $\phi$ meson induces a $\Lambda$ repulsive force and hardens the equation of state. 
The hardening arising from the $\phi$ meson compensates the softening from the existence of hyperons.
The flavor SU(3) and spin-flavor SU(6) relations are examined as well.
We found that the coupling constants fitted by neutron matter properties could yield high enough 
maximum mass of a neutron star and the obtained results satisfy both the mass and radius constraints.
The onset of the hyperon direct Urca process in neutron stars
is also investigated using our parametrization.

\end{abstract}



\date{\today}

\maketitle

\section{Introduction}
\label{sec:int}

A neutron star is a massive and dense stellar object whose major constituents are neutrons.
Due to the fermionic nature of neutrons, protons and electrons are also allowed to exist
under the condition of charge neutrality.
It was widely accepted that the canonical mass of a neutron star is about $1.4\,M_{\odot}$, 
where $M_{\odot}$ is the solar mass, and the radius is around 10~km~\cite{LP04b,Lattimer13}.
But the recent measurements suggested that the upper limit of the neutron star mass could be 
larger than $2\,M_\odot$~\cite{DPRRH10, AFWT13}.
Such a compact object would have high nuclear matter density up to several times the nuclear 
saturation density ($n_0 = 0.16 ~\si{fm}^{-3}$) in the core. 
In such a dense matter, it is natural to expect that non-nucleonic degrees of freedom may come out
and various equations of state for a neutron star have been suggested.
For instance, since the average distance between nucleons could be smaller than the de Brogile 
wave length of quarks, there may exist quark matter in the core of a neutron 
star~\cite{Weber04, ABPR04,WSPHS11}.
On the other hand, because of the bosonic nature of mesons, pion condensation~\cite{Baym73,AB74} 
and kaon condensation~\cite{TPL93,GS98c,LKHL13} were suggested to be formed in the core of 
a neutron star as examples of Bose-Einstein condensation. 
In principle, such ideas could be tested by future observations, such as
X-ray observation in low-mass X-ray binaries and gravitational 
wave detections from neutron star binary mergers.

Concerning the hadronic degrees of freedom, hadrons with strangeness are likely to be formed because
hyperon matter is energetically favorable than pure nucleon matter. 
In general, the presence of hyperons makes the equation of state (EOS) softer than that
without hyperon degrees of freedom, which reduces the maximum mass of neutron stars.
Therefore, softening of the EOS might be incompatible with the recent observation of 
$2.0~M_\odot$ neutron stars, namely, PSR J1614-2230 and PSR J0348+0432~\cite{DPRRH10, AFWT13}.
The question on the existence of hyperons in the core of neutron star is paraphrased
as the `hyperon puzzle' and understanding the structure of a neutron star with strangeness degree of freedom 
now becomes one of major research topics in neutron star physics.
For example, three-body repulsive forces among baryons was suggested as a mechanism to 
compensate the softening of EOS in Refs.~\cite{VLPPB10,YFYR13,YFYR14,LLGP14}.
In the presence of hyperons in nuclear matter, the softening of EOS may be compensated by $YNN$, $YYN$, and 
$YYY$ interactions, where $Y$ stands for a hyperon, besides the three-body $NNN$ interaction.
The authors of Refs.~\cite{YFYR13,YFYR14} proposed the multi-Pomeron exchange potential to obtain
three-body repulsive force on top of the soft core interactions for having large neutron star mass.
Based on quantum Monte Carlo simulations it was claimed that the phenomenological $\Lambda NN$ 
potential can make the threshold density for hyperon population much higher than that of the neutron star core~\cite{LLGP14}.
Hyperon puzzle may also be evaded by the onset of quark matter in the compact star.
The transition from massive hadron stars to strange quark stars suggests
two coexisting families of compact stars, and the strangeness of hyperons
is dissolved into the deconfined quark matter~\cite{BD00, BBDFL02, BPV04}.

The investigation on the role of hyperons to neutron star mass and radii has a long history
starting 1960s~\cite{AS60,Pandharipande71,BJ74,Glendenning82,GM91}.
Most of theoretical investigations are based on the relativistic mean field (RMF) 
approach~\cite{BG04,BM09b,BHZBM11,WCS12a,WCS12b,GHK14,CV15}.%
\footnote{One may extend the non-relativistic approach of the Skyrme force model by
including hyperons~\cite{BG97,LY97,GDS12} for studying hypernuclei. This approach can also
be applied to explore the neutron star structure as discussed in Refs.~\cite{Mornas04,LHKL14}.}
Due to the limitations in the observations, however, mainly neutron star mass has been 
studied and the research focus was on the change of the neutron star mass.
Throughout extensive studies, both relativistic and non-relativistic approaches seem to
find conditions to fulfill the observed maximum mass constraint of neutron 
stars~\cite{BHZBM11,WCS12a,WCS12b,CV15,Mornas04,LHKL14}.
From the recent analyses of the X-ray burst from neutron star binaries classified as low-mass X-ray 
binaries~(LMXB), both masses and radii have been estimated with significant statistics even though 
it is too early to make any firm conclusions~\cite{OGP08,SLB10,SPRW11,GSWR13}.
Firstly, a small radii of neutron stars were reported by \"{O}zel et al.~\cite{OGP08} and by 
Guillot et al.~\cite{GSWR13} from the analysis of X-ray bursts in LMXB.
By neglecting the mass dependence, Guillot et al.~\cite{GSWR13} estimated the radius of neutron 
stars to be $R_{\rm NS} = 9.1_{-1.5}^{+1.3}$~\si{km} with 90\% confidence.%
\footnote{
A recent estimate on the neutron star radius for a mass of $1.5~M_\odot$ 
gives $10.1~\si{km} \le R_{\rm NS} \le 11.1~\si{km}$~\cite{OPGBHS16}.
We also note that the causality limits the maximum neutron star mass to 
be smaller than $2.1~M_\odot$ if the radius is 9.1~km~\cite{LP06}.}

On the other hand, by considering the gravitational redshift of X-rays generated in the photosphere 
of the X-ray burst sources, Steiner et al.~\cite{SLB10} suggested that the radius of neutron stars would 
be $R_{\rm NS} = 12^{+0.5}_{-1.0}$~\si{km}.
This estimate is consistent with the experimental and theoretical studies of nuclear matter as described in
Ref.~\cite{LL12}.

From the analysis of cooling phases of an X-ray burst source, Sulumeinov et al.~\cite{SPRW11} 
suggested a stiff EOS that allows $2.3 \, M_{\odot}$ with radius larger than $14$~\si{km}. 
However, such an EOS carries a large density slope of nuclear
symmetry energy, and thus the direct Urca process may
occur for neutron stars with masses less than $1.2\,M_{\odot}$.
Hence, such models may be not easy to explain the cooling curves of neutron stars~\cite{LHL15}.
Near future, new X-ray telescope NICER (Neutron star Interior Composition ExploreR)
will be able to provide more reliable data on both masses and radii of neutron stars~\cite{OPAMB15}.
The validity of various EOS in the literature can then be tested by these new observations.

The purpose of the present work is to find an EOS in the relativistic mean field approach
which satisfies both constraints on mass and radius simultaneously with the existence of 
strangeness in a neutron star.
For this end, we first set up an SU(2) model that includes only nucleons as baryons as well 
as the $\sigma$, $\omega$, and $\rho$ mesons, and determine
the model parameters to reproduce the properties of symmetric nuclear matter and pure 
neutron matter that were obtained in Refs.~\cite{GCR11, DSS13}.
The criteria for SU(2) models was discussed in many different approaches, such as Monte Carlo 
calculations~\cite{GCR11} and chiral effective field theories~\cite{DSS13,HK16}.
The effects of three nucleon forces are also discussed in Refs.~\cite{RRH15,SLB15}.
After developing the SU(2) model, we extend it to the case of flavor SU(3) and investigate 
the role of hyperons in the mass-radius relationship of neutron stars.
For this purpose, we adopt several cases for the determination of coupling constants among hyperons
and use the estimation of Steiner et al.~\cite{SLB10} for the radii of neutron stars.
Then the obtained critical density or neutron star mass are investigated for the hyperon direct Urca process, 
which plays an important role in the thermal evolution of neutron stars in the presence of hyperons.

This paper is organized as follows. 
We introduce the Lagrangian density of nuclear matter in the flavor SU(2) model in Sec.~\ref{sec:nuc}. 
The equations of motion, energy density, and pressure are obtained and discussed in detail.
In Sec.~\ref{sec:hyper}, we extend our model to the flavor SU(3) to include strangeness 
degree of freedom for addressing the hyperon puzzle in the RMF approach.
We explain how we control the coupling constants of the model and discuss the modified equations 
of motion in hyperon matter.
In Sec.~\ref{sec:nshyper}, the numerical results for the mass-radius relation of neutron stars are summarized.
The role of hyperons in direct Urca process is then discussed in Sec.~\ref{sec:urca}.
Section~\ref{sec:con} summarizes our conclusions.
The details on how we fix the SU(2) parameters to reproduce the properties
of symmetric nuclear matter and those of pure neutron matter are 
described in Appendix.

\section{Nuclear Model}
\label{sec:nuc}

In order to describe nuclear matter, we start with the RMF model in the flavor SU(2) sector, 
which includes the $\rho$ and $\omega$ vector mesons in addition to the scalar $\sigma$ meson.
The effective Lagrangian that includes nonlinear self-interactions of the meson fields reads~\cite{SPLE04}
\begin{eqnarray}
\mathcal{L}_{\sigma\omega\rho} & = &
\sum_{B=n, p}
\bar{\psi}_B^{} \Bigl[ (i\slashed{\partial} - g_{\omega B}^{} \gamma_\mu \omega^\mu)
-  g_{\rho B}^{} \gamma_\mu \vec{\rho}^{\,\mu} \cdot \vec{\tau}_B^{}  
\nonumber \\[-10pt] && \qquad\qquad \mbox{}
- (M_B - g_{\sigma B}^{} \sigma)
- \frac{e}{2} \left(1 + \tau_3^{} \right) A_\mu \gamma^\mu \Bigr] \psi_B^{} 
\nonumber \\ && \mbox{}
+ \frac{1}{2} \partial_\mu \sigma \partial^\mu \sigma
- \frac{1}{2} m_\sigma^2 \,\sigma^2
- \frac{\kappa}{3!} \left( g_\sigma^{} \sigma \right)^3 - \frac{\lambda}{4!} \left( g_\sigma^{} \sigma \right)^4
\nonumber \\ && \mbox{}
+\frac{1}{2} m_\omega^2\, \omega_\mu \omega^\mu 
- \frac{1}{4} \omega_{\mu\nu} \omega^{\mu\nu}
+ \frac{\zeta}{4!} g_\omega^4 \left( \omega_\mu \omega^\mu \right)^2  
\nonumber \\ && \mbox{}
+ \frac{1}{2} m_\rho^2 \, \vec{\rho}_\mu \cdot \vec{\rho}^{\,\mu}
-\frac{1}{4} \vec{\rho}_{\mu\nu} \cdot \vec{\rho}^{\,\mu\nu}
+ \frac{\xi}{4!} g_\rho^4 \left( \vec{\rho}_{\mu} \cdot \vec{\rho}^{\,\mu} \right)^2 
 \nonumber \\ && \mbox{} 
+ f(\sigma,\omega_\mu \omega^\mu) g_\rho^2 \left( \vec{\rho}_{\mu} \cdot \vec{\rho}^{\,\mu} \right)
- \frac{1}{4} F_{\mu\nu} F^{\mu\nu} 
\nonumber \\ && \mbox{}
+\sum_{l=e^-,\mu^-} \bar{\psi}_l^{} \left( i\slashed{\partial} - m_l^{} \right) \psi_l^{} ,
\label{eq:lag}
\end{eqnarray}
where $M_B$ is the baryon mass,%
\footnote{
Since we are assuming SU(2) isospin symmetry, the proton and neutron have the same mass $M_N$.}
$m_l^{}$ is the lepton mass, $\vec{\tau}_B^{}$ are the Pauli matrices for isospin, and 
the field strength tensors of the photon ($A_\mu$) and vector mesons ($\omega_\mu$ and $\vec{\rho}_\mu$)
are given by
\begin{equation}
\begin{split}
F_{\mu\nu} & = \partial_\mu A_\nu -\partial_\nu A_\mu ,\\
\omega_{\mu\nu} & =  \partial_\mu\omega_\nu -\partial_\nu \omega_\mu ,\\
\vec{\rho}_{\mu\nu} & = \partial_\mu\vec{\rho}_\nu - \partial_\nu \vec{\rho}_\mu  .
\end{split}
\end{equation}
The masses of the $\sigma$, $\omega$, and $\rho$ mesons are taken as
$m_\sigma^{} = 491.5 \mbox{ MeV}$~\cite{FHPS10}, $m_\omega^{} = 782.5 \mbox{ MeV}$, and
$m_\rho^{} = 775.3  \mbox{ MeV}$~\cite{PDG16}.%
\footnote{The scalar $\sigma$ meson corresponds to the $f_0(500)$ in the list of Particle Data Group~\cite{PDG16}.
The range of the $\sigma$ meson mass is wide and we adopt the value of $m_\sigma^{}$ from Ref.~\cite{FHPS10}.}
The interactions between mesons are assumed to be
\begin{equation}
f(\sigma, \omega^\mu\omega_\mu)
= \sum_{i=1}^{6} \Lambda_{s, i} \left( g_\sigma^{} \sigma \right)^{i}
+  \sum_{i=1}^{3} \Lambda_{v, i} \left( g_\omega^2 \omega_\mu \omega^\mu \right)^{i} .
\end{equation}
The parameters $\Lambda_{s, i}$ and $\Lambda_{v, i}$ will be determined by the neutron matter 
properties~\cite{SPLE04}.
For simplicity, we introduce brief notations at the mean field level as
\begin{equation}
S = g_\sigma^{} \braket{\sigma}, \quad 
W = g_\omega^{} \braket{\omega^0}, \quad
R = g_\rho^{} \braket{\rho_z^{0}},
\end{equation}
where $\rho_z^0$ represents the time component of the $\rho$ meson with the third-component of
isospin $I_z = 0$.
Then the equations of motion for meson fields in uniform nuclear matter are obtained as
\begin{equation}
\begin{aligned}
n_s = & \left( \frac{m_\sigma^{}}{g_\sigma^{}} \right)^2 S 
+ \frac{\kappa}{2}S^2 + \frac{\lambda}{6}S^3
- R^2 \frac{\partial f}{\partial S}, \\
n = & \left(\frac{m_\omega^{}}{g_\omega^{}}\right)^2 W
+ \frac{\zeta}{6}W^3
+  R^2 \frac{\partial f}{\partial W} , \\
- \frac{1}{2}\alpha
= & \left( \frac{m_\rho^{}}{g_\rho^{}} \right)^2 R
+ \frac{\xi}{6}R^3
+ 2R f ,
\end{aligned}
\end{equation}
where $n_s$ is the total baryon scalar number density obtained as
\begin{equation}
n_s^{} = \sum_{B=n,p} \frac{1}{\pi^2}\int_{0}^{k_F^B}
\frac{dk\, k^2 M_N^*}{\sqrt{k^2 + M_N^{*2}}} ,
\end{equation}
and $n = n_n^{} + n_p^{}$ is the total baryon number density while $\alpha = n_n^{} - n_p^{}$ with
\begin{eqnarray}
n_n^{} & = & \frac{(k_F^n)^3}{3\pi^2}, \qquad
n_p^{} = \frac{(k_F^p)^3 }{3\pi^2},
\end{eqnarray}
Here $M_N^*$ is the Dirac effective mass of the nucleon defined as 
$M_N^* = M_N - g_\sigma^{} \braket{\sigma} = M_N - S$.

The energy density and pressure of the nucleonic contribution can be obtained
from the given Lagrangian and are written as
\begin{eqnarray}
\mathcal{E}_{\sigma\omega\rho} &=& \sum_{B=n,p}  \frac{1}{\pi^2}\int_{0}^{k_F^B}
dk\, k^2\sqrt{k^2 + M_N^{*2}}  
\nonumber \\ && \mbox{} 
+ \frac{1}{2} \left( \frac{m_\sigma^{}}{g_\sigma^{}} \right)^2 S^2
 +  \frac{\kappa}{3!} S^3 + \frac{\lambda}{4!} S^4 
\nonumber \\ && \mbox{}
 + \frac{1}{2} \left( \frac{m_\omega^{}}{g_\omega^{}} \right)^2 W^2  
 + \frac{\zeta}{8} W^4 
\nonumber \\ &&
+ \frac{1}{2} \left( \frac{m_\rho^{}}{g_\rho^{}} \right)^2 R^2
+ \frac{\xi}{8} R^4 
+ R^2 \left( f + W \frac{\partial f}{\partial W} \right) ,
\label{eq:ene}  \\
P_{\sigma\omega\rho} &=& \sum_{B=n,p}  \frac{1}{3\pi^2}\int_{0}^{k_F^B}
dk\, \frac{k^4}{\sqrt{k^2 + M_N^{*2}}}  
\nonumber \\ && \mbox{}
- \frac{1}{2} \left( \frac{m_\sigma}{g_\sigma} \right)^2 S^2
- \frac{\kappa}{3!} S^3 - \frac{\lambda}{4!} S^4 
 + \frac{1}{2} \left( \frac{m_\omega^{}}{g_\omega^{}}\right)^2 W^2 
 \nonumber \\ && \mbox{} 
 + \frac{\zeta}{4!} W^4 + \frac{1}{2} \left( \frac{m_\rho^{}}{g_\rho^{}} \right)^2 R^2
+ \frac{\xi}{4!} R^4 + f R^2  .
\label{eq:pre}
\end{eqnarray}
The chemical potentials of neutrons and protons, which are the eigenvalues of the
Fermi energies $k_F^n$ and $k_F^p$, respectively, are obtained as
\begin{equation}
\begin{aligned}
\mu_n^{} = & \sqrt{(k_F^n)^2 + M_N^{*2}} + W - \frac{1}{2}R\,, \\
\mu_p^{} = & \sqrt{(k_F^p)^2 + M_N^{*2}} + W + \frac{1}{2}R\,.
\end{aligned}
\end{equation}

The coupling constants $g_\sigma^{}$, $g_\omega^{}$, $\kappa$, and $\lambda$
are determined to reproduce the properties of the symmetric nuclear matter.
In the present calculation, we use the nuclear saturation density $n_0^{}$ and
\begin{equation}
\begin{aligned}
& M_N^{*}  = 0.75\, M_N, \qquad
B = 16 \mbox{ MeV}, \\
& 
K = 240\mbox{--}245 \mbox{ MeV},
\end{aligned}
\label{eq:matter_property}
\end{equation}
where $B$ is the binding energy per nucleon and $K$ is the incompressibility coefficient.
In this fitting procedure, we fix $\zeta = 0$ so that the rescaling of $g_{\omega}^{}$ can be used
even in the presence of the $\phi$ meson.
This will be discussed in the next section.

The remaining parameters $\xi$, $g_\rho^{}$, $\Lambda_{s,i}$, and $\Lambda_{v,i}$
are determined by the neutron matter properties reported in Refs.~\cite{GCR11,DSS13}. 
This is done by using the polynomial parametrization for the energy per baryon
in neutron matter suggested in Ref.~\cite{GISPF09}, which reads
\begin{equation}
\frac{E}{A} = a \left( \frac{n}{n_0} \right)^{\alpha} 
+ b \left( \frac{n}{n_0} \right)^{\beta}.
\end{equation}
In Ref.~\cite{GCR11}, Gandolfi, Carlson, and Reddy (GCR) used quantum Monte Carlo techniques 
to estimate the equation of state of neutron matter, while Drischler, Soma, and Schwenk (DSS)
used chiral effective field theory~\cite{DSS13}.
For our numerical calculations, we use GCR5 and DSS2 parameterizations for neutron matter as
compiled in Table~1 of Ref.~\cite{SLB15}.
Explicitly, we adopt
\begin{eqnarray}
\begin{aligned}
& a = 13.0 \mbox{ MeV}, \quad \alpha = 0.50, \\
& b = 4.71 \mbox{ MeV}, \quad \beta = 2.49,
\end{aligned}
\label{NM:GCR}
\end{eqnarray}
for GCR5 and
\begin{eqnarray}
\begin{aligned}
& a = 11.95 \mbox{ MeV}, \quad \alpha = 0.495, \\
& b = 3.493 \mbox{ MeV}, \quad \beta = 2.632,
\end{aligned}
\label{NM:DSS}
\end{eqnarray}
for DSS2.

\begin{table*}[t]
\caption{The fitted parameter sets of SU(2) RMF models.
RGCR represents RMF model with GCR5 parametrization and RDSS represents RMF models
with DSS2 parametrization. 
For comparison, the fitted parameters in other works are also presented with references.
}
\begin{tabular}{cccccccl}
\toprule
Parameter         &    RGCR            &   RDSS           & IU-FSU~\cite{FHPS10}    
                  &   SFHo~\cite{SHF12}   & GM1~\cite{GM91}    & NL3~\cite{LKR96}   &  Unit   \\
\hline
$m_\sigma^{}$        & $ 2.491 $  
                  & $ 2.491 $  
                  & $ 2.491 $  
                  & $ 2.371 $  
                  & $ 2.491$   
                  & $ 2.575$   
                  & fm$^{-1}$ \\
$m_\omega^{}$        & $3.966 $   
                  & $3.966 $   
                  & $3.966 $   
                  & $3.864 $   
                  & $3.966 $   
                  & $3.966 $   
                  & fm$^{-1}$ \\
$m_\rho^{}$          & $3.929 $   
                  & $3.929 $   
                  & $3.867 $   
                  & $3.902 $   
                  & $3.867 $   
                  & $3.867 $   
                  & fm$^{-1}$ \\
$g_{\sigma N}^{}$     &  $8.005$   
                  &  $7.985$   
                  &  $9.971$   
                  &  $7.536$   
                  &  $8.553$   
                  &  $10.217$  
                  &   \\
$g_{\omega N}^{}$     &  $9.235$    
                  &  $9.235$    
                  &  $13.032$   
                  &  $8.782$    
                  &  $10.603$   
                  &  $12.868$   
                  &   \\
$g_{\rho N}^{}$       &  $11.108$     
                  &  $11.033$     
                  &  $13.590$     
                  &  $9.384$      
                  &  $8.121$      
                  &  $8.922$      
                  &   \\
$\kappa$          &  $6.603\times 10^{-2}$     
                  &  $6.350\times 10^{-2}$     
                  &  $1.713\times 10^{-2}$     
                  &  $7.105\times 10^{-2}$     
                  &  $2.805\times 10^{-2}$     
                  &  $1.956\times 10^{-2}$     
                  &  fm$^{-1}$ \\
$\lambda$         &   ~$-2.900\times 10^{-2}$~      
                  &   ~$-2.474\times 10^{-2}$~      
                  &   ~$2.960\times 10^{-4}$~       
                  &   ~$-2.645\times 10^{-2}$~     
                  &   ~$-6.420\times 10^{-3}$~     
                  &   ~$-1.591\times 10^{-2}$~     
                  &   \\
$\zeta$           &  $-$                           
                  &  $-$                           
                  &  $3.0\times 10^{-2}$           
                  &  $-1.701\times 10^{-3}$        
                  &  $-$                           
                  &  $-$                           
                  &   \\
$\xi$             &  $-3.807\times 10^{-5}$         
                  & $-1.088 \times 10^{-7}$         
                  &  $-$                            
                  &  $3.453\times 10^{-3}$          
                  &  $-$                            
                  &  $-$                            
                  & \\
$\Lambda_{s1}$    &  $-3.788\times 10^{-4}$         
                  & $4.467 \times 10^{-4}$          
                  &  $-$                            
                  &  $-3.054\times 10^{-2}$         
                  &   $-$                           
                  &   $-$                           
                  & fm$^{-1}$\\
$\Lambda_{s2}$    & ~$1.810\times 10^{-2}$~       
                  & ~$4.267 \times 10^{-2}$~      
                  &  $-$                          
                  &  $1.021\times 10^{-2}$        
                  &   $-$                         
                  &   $-$                         
                  &   \\
$\Lambda_{s3}$    &  $1.724\times 10^{-2}$         
                  &  $-3.597 \times 10^{-4} $      
                  &  $-$                           
                  &  $8.048\times 10^{-4}$         
                  &   $-$                          
                  &   $-$                          
                  & fm$^{\phantom{-1}}$\\  
$\Lambda_{s4}$    &  $2.424\times 10^{-3}$         
                  &  $2.550 \times 10^{-4}$        
                  &  $-$                           
                  &  $1.072\times 10^{-3}$         
                  &   $-$                          
                  &   $-$                          
                  & fm$^{2}$\\
$\Lambda_{s5}$    &  $-2.862\times 10^{-3}$        
                  &  $2.588 \times 10^{-3}$        
                  &  $-$                           
                  &  $5.542\times 10^{-5}$         
                  &   $-$                          
                  &   $-$                          
                  & fm$^{3}$\\
$\Lambda_{s6}$    &  $-3.416\times 10^{-8}$        
                  &  $9.217 \times 10^{-8}$        
                  &  $-$                           
                  &  $3.606\times 10^{-6}$         
                  &   $-$                          
                  &   $-$                          
                  & fm$^{4}$\\
$\Lambda_{v1}$    &  $1.131\times 10^{-4}$         
                  &  $2.220\times 10^{-5}$         
                  & ~$4.60\times 10^{-2}$~         
                  &  $7.616\times 10^{-2}$         
                  &   $-$                          
                  &   $-$                          
                  &  \\
$\Lambda_{v2}$    &  $-6.174\times 10^{-4}$        
                  &  $-8.536\times 10^{-5} $       
                  &  $-$                           
                  &  $-2.765\times 10^{-4}$        
                  &   $-$                          
                  &   $-$                          
                  & fm$^{2}$\\
$\Lambda_{v3}$    &  $1.563\times 10^{-5}$       
                  &  $5.560\times 10^{-6}$       
                  &  $-$                         
                  &  $6.861\times 10^{-4}$       
                  &   $-$                        
                  &   $-$                        
                  & fm$^{4}$\\
\hline
\end{tabular}\label{tb:rmfpa}
\end{table*}

With these information one can calculate the properties of nuclear and neutron matter, which
determines the coupling constants and other model parameters.
The details can be found in Appendix, and the parameter sets obtained in this way 
are presented in Table~\ref{tb:rmfpa}. 
The parameter set RGCR (i.e., the RMF model with the GCR parameterization) is obtained with the GCR5
parametrization
and the set RDSS is with the DSS2 parametrization.

\begin{figure}[t]
\includegraphics[scale=0.45]{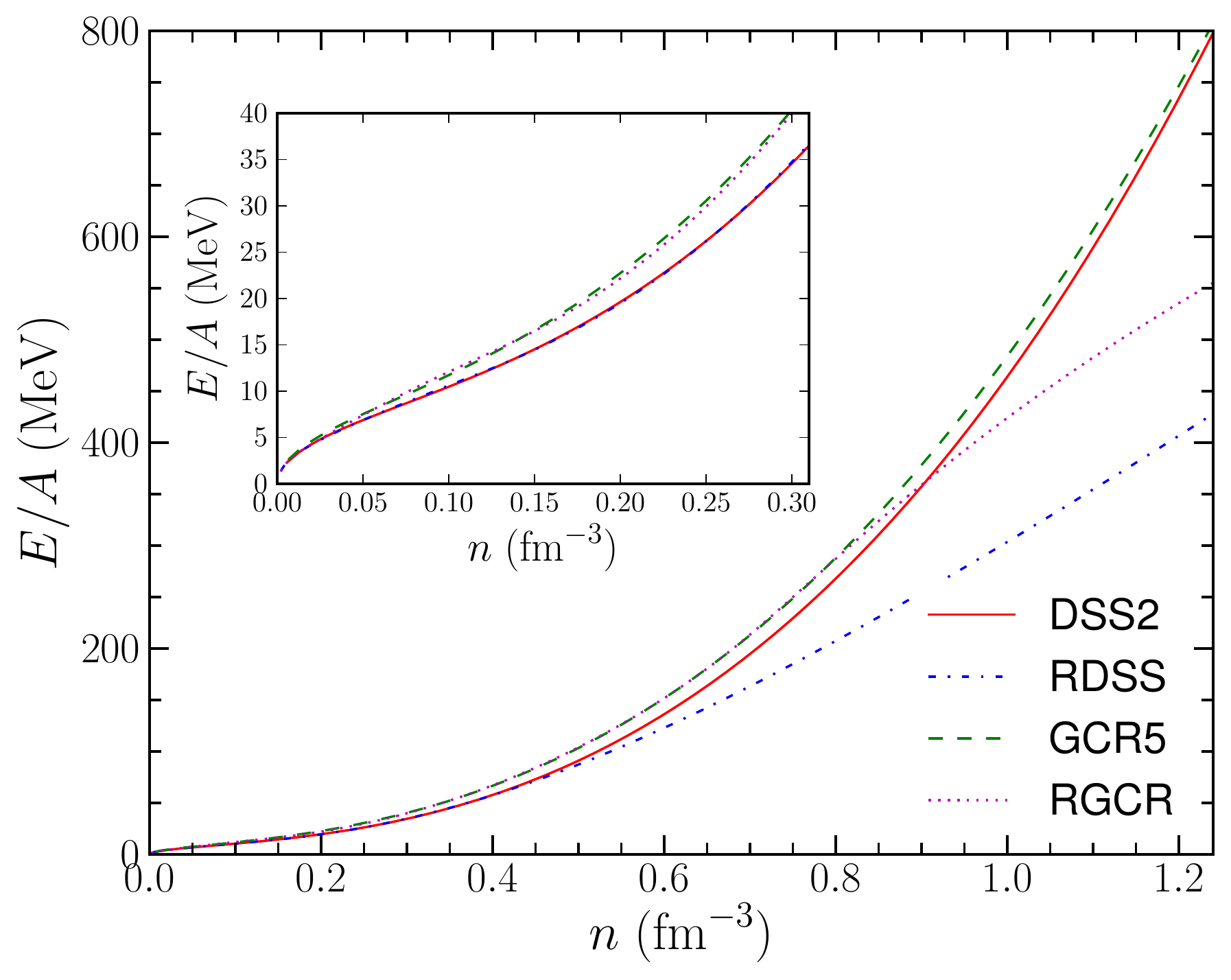}
\caption{Energy per baryon of pure neutron matter.
RGCR and RDSS are obtained from the Lagrangian of Eq.~\eqref{eq:lag} by fitting to the results of
GCR5~\cite{GCR11} and DSS2~\cite{DSS13}, respectively.
}
\label{fig:epa}
\end{figure}

For comparison, we show the energy per baryon of pure neutron matter in Fig.~\ref{fig:epa}.
The small window magnifies the results in the low density region, $0 \le n \le 0.3~\si{fm}^{-3}$.
The solid and dashed lines show the theoretical results of DSS2 and GCR5 models, respectively,
while the results of RMF with parameter sets RDSS and RGCR are presented by dot-dashed and dotted 
lines, respectively.
This shows that, for pure neutron matter, the RDSS gives consistent results with those of DSS2 up to 
$n \sim 0.5~\si{fm}^{-3} \sim 3\, n_0^{}$ and the agreement between GCR5 and RGCR goes up to 
$n \sim 0.8~\si{fm}^{-3} \sim 5\, n_0$.

\begin{table}[t]
\caption{Nuclear matter properties at the saturation density.
In our calculations (RGCR and RDSS), the saturation density $n_0^{}$, binding energy per nucleon $B$, and
the incompressibility coefficient $K$ are inputs while the symmetry energy $J$ and the symmetry energy 
slope $L$ are computed at $n_0^{}$.
For comparison, the results of other models are presented as well.}
\begin{tabular}{ccccccc}
\toprule
           & RGCR     & RDSS   & IU-FSU  & SFHo   & GM1 & NL3  \\
 \hline
 $n_0^{}$ ($\si{fm}^{-3}$)  &  ~0.160~    & ~0.160~   &  ~0.155~  & ~0.158~  &  ~0.153~  & ~0.148~  \\
 ~$B$   (MeV)~               &  16.0       &  16.0     &  16.4     & 16.2     &   16.3    & 16.2   \\
 $K$   (MeV)                 &  240        & 245       & 231       & 245      &   300     & 272         \\
 $J$ (MeV)                 &  32.9       & 30.7      & 31.3      & 31.6     &   32.5    & 37.3 \\
 $L$   (MeV)                 &  46.8       & 42.2      & 47.2      & 52.9     &   94.0    & 118\\
 \hline
 Reference & & & \cite{FHPS10} & \cite{SHF12} & \cite{GM91} & \cite{LKR96} \\ \hline
\end{tabular}\label{tb:rmf_sym}
\end{table}

Table~\ref{tb:rmf_sym} shows the standard nuclear matter properties for various models
considered in this work.
In the present calculation, $n_0^{}$, $B$, and $K$ are inputs and $J$ and $L$ are calculated results.
For comparison, the corresponding quantities of other models in the literature are given as well.
These SU(2) models are then used to compute the neutron star mass.
As will be discussed in Sec.~\ref{sec:nshyper}, these models can give large neutron star masses.
On top of these models, we introduce hyperon degrees of freedom in the next Section.

\section{Model for hyperon matter}
\label{sec:hyper}

In order to study the role of strangeness degree of freedom in neutron star structure, 
we extend the Lagrangian of Eq.~\eqref{eq:lag} by introducing the spin-1/2 flavor-octet hyperons.
In addition, to study the role of the scalar and vector mesons with hidden strangeness, 
we include the $\phi(1020)$ vector meson and the $f_0(980)$ scalar meson. 
Therefore, we have the SU(3)$_F$ nonet structure for scalar and vector mesons~\cite{SDGGS93,BM09b}.
In the following we denote the $f_0(980)$ meson field by $\sigma^*$.
Then the Lagrangian density reads
\begin{equation}
\mathcal{L} =
\mathcal{L}_{\sigma\omega\rho}^\prime + \mathcal{L}_{\sigma^* \phi},
\label{eq:lag_hy}
\end{equation}
where $\mathcal{L}_{\sigma\omega\rho}^\prime$ has the same form as $\mathcal{L}_{\sigma\omega\rho}$ 
in Eq.~\eqref{eq:lag} but includes hyperon octet such that $B = p, n, \Lambda, \Sigma^{\pm,0}, \Xi^{0,-}$, 
and $\mathcal{L}_{\sigma^* \phi}$ includes the terms concerning the $\sigma^*$ and the $\phi$.
Explicitly, it reads
\begin{eqnarray}
\mathcal{L}_{\sigma^* \phi} &=& 
\sum_B \bar{\psi}_B^{} \left( g_{\sigma^* B}^{} \sigma^* -  g_{\phi B}^{} \gamma_\mu \phi^\mu \right) \psi_B^{}
\nonumber \\ && \mbox{} 
+ \frac{1}{2}\partial_\mu \sigma^* \partial^\mu \sigma^* 
- \frac{1}{2}m_{\sigma^*}^2\, \sigma^{*2} 
\nonumber \\ && \mbox{} 
+ \frac{1}{2} m_\phi^2 \, \phi_\mu \phi^\mu 
- \frac{1}{4} \phi_{\mu\nu} \phi^{\mu\nu} ,
\end{eqnarray}
where $m_{\sigma^*}^{}$ and $m_\phi^{}$ are the $f_0(980)$ and $\phi(1020)$ masses and we use 
$m_{\sigma^*}^{} = 975$~MeV and $m_\phi^{} = 1020$~MeV.
For hyperon masses, we also use the values provided by the Particle Data Group~\cite{PDG16}.
The field strength tensor of the $\phi$ meson is
\begin{equation}
\phi_{\mu\nu}  = \partial_\mu\phi_\nu - \partial_\nu \phi_\mu \,.\\
\end{equation}

Then the equations of motion of the meson fields in uniform matter can be obtained as
\begin{align}
& \sum_{B} x_{\sigma B}^{} \, n_B^{S} 
=  \left( \frac{m_\sigma}{g_{\sigma N}^{}} \right)^2 S 
+ \frac{\kappa}{2} S^2 + \frac{\lambda}{6} S^3
- R^2 \frac{\partial f}{\partial S}\,, \label{eq:ysig}\\
& \sum_{B} x_{\omega B}^{} \, n_B^B  =  \left( \frac{m_\omega^{}}{g_{\omega N}^{}} \right)^2 W
+ \frac{\zeta}{6} W^3 +  R^2 \frac{\partial f}{\partial W}\,, \label{eq:yome} \\
& \sum_{B} x_{\rho B}^{} \, n_B^B\, \tau_{3B}^{}
 =  \left( \frac{m_\rho}{g_{\rho N}^{}} \right)^2 R + \frac{\xi}{6} R^3 + 2R f \,, \label{eq:yrho}\\
& \sum_{B} x_{\sigma^* B}^{} \, n_B^S 
= \left( \frac{m_{\sigma^* }}{g_{\sigma^* \Lambda}} \right)^2 S^* , 
\label{eq:ysigst} \\
& \sum_{B} x_{\phi B}^{} \, n_B^B 
= \left(\frac{m_{\phi}^{}}{g_{\phi \Lambda}^{}} \right)^2 \Phi \,,
\label{eq:yphi}
 \end{align}
where $\Phi = g_{\phi \Lambda}^{} \braket{\phi}$ and $S^{*} = g_{\sigma^* \Lambda}^{} \braket{\sigma^*}$.
Here, $\tau_{3B}^{}$ is the $z$-component of the isospin quantum number of the baryon $B$
and our conventions are
\begin{eqnarray}
&& \tau_{3p}^{} = +\frac12, \quad \tau_{3n}^{} = - \frac12, \quad \tau_{3\Lambda}^{} = 0, 
\nonumber \\ &&
\tau_{3\Sigma^+}^{} = +1, \quad 
\tau_{3\Sigma^0}^{} = 0, \quad \tau_{3\Sigma^-}^{} = -1,
\nonumber \\ && 
\tau_{3\Xi^0}^{} = +\frac12, \quad \tau_{3\Xi^-}^{} = - \frac12.
\end{eqnarray}
The ratios of coupling constants are defined by
\begin{equation}
\begin{split}
& x_{\sigma B}^{} = \frac{g_{\sigma B}^{}}{g_{\sigma N}^{}}, \quad
x_{\omega B}^{} = \frac{g_{\omega B}^{}}{g_{\omega N}^{}}, \quad
x_{\rho B}^{} = \frac{g_{\rho B}^{}}{g_{\rho N}^{}}, \\
& x_{\sigma^* B}^{} = \frac{g_{\sigma^* B}^{}}{g_{\sigma^* \Lambda}^{}},
\quad
x_{\phi B}^{} = \frac{g_{\phi B}^{}}{g_{\phi \Lambda}^{}}\,.
\end{split}
\end{equation}
The scalar and vector densities are 
\begin{equation}
\begin{split}
n_B^S & = \braket{ \bar{\psi}_B^{} \psi_B^{} } =
           \frac{2J_B + 1}{(2\pi)^3}
\int_{0}^{k_F^B} d^3k \frac{M_{B}^{*}}{\sqrt{k^2 + M_{B}^{*2}}}\,,\\
n_B^B & = \braket{ \psi_B^{\dagger} \psi_B^{} } =
         \frac{2J_B + 1}{(2\pi)^3}\int_{0}^{k_F^B} d^3k \,,
\end{split}
\end{equation}
where $J_B$ is the spin of the baryon and $M_B^*$ is the effective mass defined as
\begin{equation}
M_B^* = M_B - x_{\sigma B}^{} S - x_{\sigma^* B}^{} S^* .
\end{equation}

In this approach, the chemical potentials or the Fermi energies of baryons and leptons are
given as
\begin{eqnarray}
\mu_B^{} &=& x_{\omega B}^{} W + x_{\rho B}^{} \tau_{3B}^{} R
+ x_{\phi B}^{} \Phi + \sqrt{(k_F^{B})^2 + (M_{B}^*)^2},
\nonumber \\
\mu_l^{} &=& \sqrt{(k_F^{l})^2 + m_l^2}. 
\end{eqnarray}
The energy density and pressure are the same as in Eqs.~\eqref{eq:ene} and \eqref{eq:pre} 
except that the contributions from the $\sigma^*$ and $\phi$ mesons are included and the 
summation is now over all hyperons.
We then obtain
\begin{eqnarray}
\mathcal{E} &=& \mathcal{E}_{\sigma \omega \rho}^\prime
+ \frac{1}{2} \left(\frac{m_{\sigma^*}^{}}{g_{\sigma^* \Lambda}^{}}\right)^2 S^{*2}
+ \frac{1}{2} \left(\frac{m_{\phi}^{}}{g_{\phi \Lambda}^{}}\right)^2 \Phi^2 ,\\
P &=&  P_{\sigma \omega \rho}^\prime
   - \frac{1}{2} \left(\frac{m_{\sigma^*}^{}}{g_{\sigma^* \Lambda}^{}}\right)^2 S^{*2}
+ \frac{1}{2} \left(\frac{m_{\phi}^{}}{g_{\phi \Lambda}^{}}\right)^2 \Phi^2 ,
\end{eqnarray}
where the prime in $\mathcal{E}_{\sigma \omega \rho}$ and $P_{\sigma \omega \rho}$
means that the summation over baryons is extended to include hyperons.

\subsection{Couplings with vector mesons}

The coupling constants between baryons and mesons determine the strength of interactions
and thus affect the EOS of hyperon matter.
Following the previous works, we make use of the flavor SU(3) and spin-flavor SU(6) relations.%
\footnote{See, for example, Refs.~\cite{DG85,SDGGS93,SDGGS94,WCS12b}. The general group
structure of meson-baryon interactions can be found, for example, in Ref.~\cite{OK04}.}
Starting from the flavor structure, the interaction Lagrangian of baryons and mesons can be 
constructed as follows.
First, the SU(3) baryon octet can be written in a matrix form as
\begin{equation}
B = \left( \begin{array}{ccc} 
\frac{1}{\sqrt2} \Sigma^0 + \frac{1}{\sqrt6} \Lambda^0 & \Sigma^+ & p \\
\Sigma^- & - \frac{1}{\sqrt2} \Sigma^0 + \frac{1}{\sqrt6} \Lambda^0 & n \\
- \Xi^- & \Xi^0 & - \sqrt{\frac{2}{3}} \Lambda^0 \end{array} \right).
\end{equation}
Similarly, the vector meson octet can be written as
\begin{equation}
V_8 = \left( \begin{array}{ccc} 
\frac{1}{\sqrt2} \rho^0 + \frac{1}{\sqrt6} \omega_8^{} & \rho^+ & K^{*+} \\
\rho^- & - \frac{1}{\sqrt2} \rho^0 + \frac{1}{\sqrt6} \omega_8^{} & K^{*0} \\
K^{*-} & \bar{K}^{*0} & - \sqrt{\frac{2}{3}} \omega_8^{} \end{array} \right)
\end{equation}
and the vector meson singlet takes the simple form of 
\begin{equation}
V_1 = \frac{\omega_1^{}}{\sqrt3} \left( \begin{array}{ccc} 1 & 0 & 0 \\ 0 & 1 & 0 \\ 0 & 0 & 1 
\end{array} \right).
\end{equation}
Then the flavor SU(3) invariant interactions between baryon octet and meson octet 
can be written as
\begin{eqnarray}
\mathcal{L}_{V_8BB} &=& \sqrt{2} g_8^{} \bigl\{ (d+f) \mbox{Tr} \left( \bar{B} B V_8 \right)
\nonumber \\ &&  \mbox{} \quad \quad 
+ (d-f) \mbox{Tr} \left( \bar{B} V_8 B \right) \bigr\},
\end{eqnarray}
while we have
\begin{eqnarray}
\mathcal{L}_{V_1BB} = \sqrt{3} g_1^{} \mbox{Tr} \left(V_1  \bar{B} B \right) 
\end{eqnarray}
for the interactions between baryon octet and meson singlet.
Therefore, we have three parameters, $g_1^{}$, $g_8^{}$, and $\alpha_V^{} \equiv f/(f+d)$,
with the condition that $f+d=1$, to completely determine the coupling constants.%
\footnote{
In the present work, we do not include the $K^*$ vector meson, even though the $K^*$ can give an extra repulsion, 
because it should be treated with kaons in the medium. This may cause kaon condensation in the core 
of neutron stars, which is, however, beyond the scope of this work.}
The physical $\omega$ and $\phi$ mesons are combinations of $\omega_8^{}$ and $\omega_1^{}$,
whose quark contents are 
\begin{eqnarray}
\ket{\omega_8^{}} &=& \frac{1}{\sqrt6} \left( \ket{\bar{u}u} + \ket{\bar{d}d} -2 \ket{\bar{s}s} \right), 
\nonumber \\
\ket{\omega_1^{}} &=& \frac{1}{\sqrt3} \left( \ket{\bar{u}u} + \ket{\bar{d}d} + \ket{\bar{s}s} \right).
\end{eqnarray}
By introducing the mixing angle $\theta$ the physical states are constructed as
\begin{eqnarray}
\ket{\phi} &=& \cos\theta \ket{\omega_8^{}} - \sin\theta  \ket{\omega_1^{}} ,
\nonumber \\ 
\ket{\omega} &=& \sin\theta \ket{\omega_8^{}} + \cos\theta  \ket{\omega_1^{}} .
\end{eqnarray}
Throughout the present work, we assume the ideal mixing, $\cos\theta = \sqrt{2/3}$, between the octet 
and singlet mesons, so that the $\phi$ is a pure $s\bar s$ state and the $\omega$ does not contain the
hidden strangeness component.

Then by introducing $z = g_8^{}/g_1^{}$, one could obtain the following relations for 
coupling constants:
\begin{eqnarray}
g_{\omega N}^{} &=& \left\{ \sqrt{\frac23} - \frac13 \left( 1 - 4 \alpha_V^{} \right) z \right\} g_1^{} ,
\nonumber \\
g_{\phi N}^{} &=& \left\{ -\frac{1}{\sqrt3} - \frac{\sqrt2}{3} \left( 1 - 4 \alpha_V^{} \right) z \right\} g_1^{} ,
\nonumber \\
g_{\omega \Lambda}^{} &=& \left\{ \sqrt{\frac23} - \frac23 \left( 1 - \alpha_V^{} \right) z \right\} g_1^{} ,
\nonumber \\
g_{\phi \Lambda}^{} &=& \left\{ -\frac{1}{\sqrt3}  - \frac{2\sqrt2}{3} \left( 1 - \alpha_V^{} \right) z \right\} g_1^{} ,
\nonumber \\
g_{\omega \Sigma}^{} &=& \left\{ \sqrt{\frac23}  + \frac23 \left( 1 - \alpha_V^{} \right) z \right\} g_1^{} ,
\nonumber \\
g_{\phi \Sigma}^{} &=& \left\{ -\frac{1}{\sqrt3}  + \frac{2\sqrt2}{3} \left( 1 - \alpha_V^{} \right) z \right\} g_1^{} ,
\nonumber \\
g_{\omega \Xi}^{} &=& \left\{ \sqrt{\frac23}  - \frac13 \left( 1 + 2 \alpha_V^{} \right) z \right\} g_1^{} ,
\nonumber \\
g_{\phi \Xi}^{} &=& \left\{ -\frac{1}{\sqrt3}  - \frac{\sqrt2}{3} \left( 1 + 2\alpha_V^{} \right) z \right\} g_1^{} 
\end{eqnarray}
and
\begin{eqnarray}
&& g_{\rho N}^{} = z g_1^{}, 
\quad
g_{\rho \Sigma}^{} = 2 \alpha_V^{} z g_1^{},
\nonumber \\
&& g_{\rho \Xi}^{} = - \left( 1 - 2 \alpha_V^{} \right) z g_1^{}. 
\end{eqnarray}
Note that $g_{\rho \Lambda} = 0$ because of isospin symmetry.

The coupling constants are constrained to some extent at free space.
For example, $\alpha_V^{} = 1$ is favored in the Nijmegen soft core potential~\cite{RSY99}
and in the QCD sum rule analysis of Ref.~\cite{ETR06}.
However, it is not yet clear how these couplings would change in nuclear medium, in particular,
in a high density region like inside neutron stars, which will eventually affect the properties of neutron stars.
The purpose of the present work is, therefore, to see whether one can satisfy the mass-radius constraint of neutron stars by
varying the coupling constant parameters, $g_1^{}$ (or $g_8^{}$), $\alpha_V^{}$, and $z$.
This would give us another viewpoint on the hyperon puzzle.
To this end, we consider the following four cases.

\begin{description}
\item[Case I] 
In this case, we consider the SU(6) limit, where 
\begin{equation}
\alpha_V^{}=1, \qquad z= 1/\sqrt6
\end{equation}
as used in Refs.~\cite{SDGGS93,SDGGS94}.
Therefore, the only parameter is the overall scale of the coupling, namely, $g_1^{}$.
In this case, we have very strong constraints on the coupling constants as
\begin{equation}
\begin{aligned}
& \textstyle \frac{1}{3}g_{\omega N}^{}  = \frac{1}{2} g_{\omega \Lambda}^{}
= \frac{1}{2}g_{\omega \Sigma}^{} = g_{\omega\Xi}^{} ,\\
& \textstyle g_{\rho N}^{} = \frac{1}{2} g_{\rho \Sigma}^{} = g_{\rho \Xi}^{} \,,\\
& \textstyle g_{\phi \Lambda}^{} = g_{\phi \Sigma}^{} = \frac{1}{2} g_{\phi \Xi}^{}
= -\frac{\sqrt{2}}{3} g_{\omega N}^{}\,, 
\quad g_{\phi N}^{} = 0\,.
\end{aligned}
\label{eq:case1}
\end{equation}

\item[Case II] 
Here, we set $\alpha_V^{} = 1$ but vary the value of $z$.
In this case, the relations between coupling constants read
\begin{equation}
\begin{aligned}
& \frac{g_{\omega\Lambda}^{}}{g_{\omega N}^{}} = \frac{g_{\omega \Sigma}^{}}{g_{\omega N}^{}} 
= \frac{\sqrt{2}}{\sqrt{2} + \sqrt{3}z}\,,
\\
& \frac{g_{\omega\Xi}^{}}{g_{\omega N}^{}} 
= \frac{\sqrt{2} -\sqrt{3}z}{\sqrt{2} + \sqrt{3}z}\,,
\quad
\frac{g_{\phi\Lambda}^{}}{g_{\omega N}^{}} 
= \frac{g_{\phi \Sigma}^{}}{g_{\omega N}^{}} 
\frac{-1 }{\sqrt{2} + \sqrt{3}z}\,,
\\
& \frac{g_{\phi N}^{}}{g_{\omega N}^{}} 
 = \frac{\sqrt{6}z - 1}{\sqrt{2} + \sqrt{3}z}\,,
 \quad
\frac{g_{\phi\Xi}^{}}{g_{\omega N}^{}} 
= -\frac{1 + \sqrt{6}z}{\sqrt{2} + \sqrt{3}z}\,,\\
& g_{\rho N}^{} = \frac{1}{2} g_{\rho \Sigma}^{} = g_{\rho \Xi}^{} .
\end{aligned}
\end{equation}
Note that in the presence of the $\phi$ meson with non-vanishing
$g_{\phi N}$, the symmetric nuclear matter properties are modified. 
Thus it is necessary to rescale the coupling constant of $g_{\omega N}^{}$ to
preserve the nuclear matter properties~\cite{WCS12b}, which leads to%
\footnote{This rescaling method is not applicable if $\zeta \ne 0$ (see Table~\ref{tb:rmfpa}).
In the case of $\zeta \ne 0$, it is necessary to solve nonlinear equations involving
$\omega$ as in Eq.~\eqref{eq:omega}. It also requires to rescale $\zeta$ to keep the
symmetric nuclear matter properties at the saturation density. Therefore, we do not
consider the case of $\zeta \neq 0$ in the present work.}
\begin{equation}
\begin{aligned}
& g_{\omega N}^{\rm rs} = g_{\omega N}^{}
\sqrt{1 + \frac{g_{\phi N}^2}{g_{\omega N}^2} \frac{m_\omega^2}{m_\phi^2}}\,, \\
& g_{\phi N}^{\rm rs} = \frac{g_{\phi N}^{}}{g_{\omega N}^{}} g_{\omega N}^{\rm rs}\,.
\label{eq:rescale}
\end{aligned}
\end{equation}

\item[Case III] 
We now fix the value of $z$ as $z = 1/\sqrt6$ and vary the value of $\alpha_V^{}$.
The relations between coupling constants become
\begin{equation}
\begin{aligned}
& \frac{g_{\omega\Lambda}^{}}{g_{\omega N}^{}} 
= \frac{2\alpha_V^{} + 4}{4\alpha_V^{} + 5}\,,
\quad
\frac{g_{\omega\Sigma}^{}}{g_{\omega N}^{}} 
= \frac{8 -2\alpha_V^{}}{4 \alpha_V^{} +5} , \\
& \frac{g_{\omega\Xi}^{}}{g_{\omega N}^{}} 
= \frac{5 -2\alpha_V^{}}{4\alpha_V^{} + 5},
\quad
\frac{g_{\phi\Lambda}^{}}{g_{\omega N}^{}} 
 = \sqrt{2}\, \frac{2\alpha_V^{} -5}{4\alpha_V^{} + 5}, \\
& \frac{g_{\phi N}^{}}{g_{\omega N}^{}} 
= \sqrt{2}\, \frac{4\alpha_V^{} -4}{4\alpha_V^{} + 5},
\quad
\frac{g_{\phi\Sigma}^{}}{g_{\omega N}^{}} 
 = - \sqrt{2}\, \frac{2\alpha_V^{} + 1}{4\alpha_V^{} + 5}, \\
& \frac{g_{\phi\Xi}^{}}{g_{\omega N}^{}} 
= - \sqrt{2}\, \frac{2\alpha_V^{} +4}{4\alpha_V^{} + 5},\\
& \frac{g_{\rho\Sigma}^{}}{g_{\rho N}^{}} 
= 2\alpha_V^{} ,
\quad
\frac{g_{\rho\Xi}^{}}{g_{\rho N}^{}} 
 = 2\alpha_V^{} -1 .
\end{aligned}
\end{equation}

\item[Case IV]
In this case, we freely vary the values of both $\alpha_V^{}$ and $z$ without any other constraints
or assumptions.
 
\end{description}

With the couplings between vector mesons and octet baryons determined above, we examine the
effects of the coupling constants on the mass-radius relation of neutron stars by varying the two parameters,
$\alpha_V^{}$ and $z$.
To determine the overall scale of the SU(3) couplings, i.e., $g_1^{}$, we use $g_{\omega N}^{}$ so that each model
keeps the value of $g_{\omega N}^{}$ for a given values of $\alpha_V^{}$ and $z$.
This does not change the nuclear matter properties determined in the SU(2) models.

\subsection{Couplings with scalar mesons}

The flavor nature of the scalar meson nonet is not yet clearly known since it may be described as an excitation
of quark-antiquark pair or as a tetraquark~\cite{BFSS99}. 
Furthermore, the mixing angle between the scalar meson octet and singlet also depends on the flavor
structure.
In the present article, following the previous works, we assume that the scalar mesons have $q\bar{q}$
structure and are ideally mixed so that the $\sigma$ meson contains light $q\bar q$ pairs only and the $f_0(980)$ 
has the hidden strangeness ($s\bar s$) structure like the $\phi$ meson.
This then leads to
\begin{equation}
g_{\sigma^* N}^{} = 0\,, \quad g_{\sigma^* \Lambda}^{} = g_{\sigma^* \Sigma}^{} .
\end{equation}
In fact, this corresponds to the relations given in Eq.~(\ref{eq:case1}) with the replacement of the $\phi$
by the $f_0(980)$.

The coupling constants $g_{\sigma Y}^{}$ can be found from the potential depths $U_Y^{(N)}$ 
in the nucleon bath with the Hugenhlotz-Van Hove theorem~\cite{GM91}. 
The potential depth is the binding energy of a hyperon $Y$ in the bath of symmetric nuclear matter 
at saturation density and can be written as
\begin{eqnarray}
U_Y^{(N)}  &=& \Big(\frac{B}{A} \Big)_{Y} = g_{\omega Y}^{} \braket{\omega^0}  + M_Y^* - M_Y 
\nonumber \\ 
&=& g_{\omega Y}^{} \braket{\omega^0} - g_{\sigma Y}^{} \braket{\sigma}
= x_{\omega Y}^{} W - x_{\sigma Y}^{} S\,.
\end{eqnarray}
In the present calculation, we adopt $U_{\Lambda}^{(N)} = -30$~MeV, $U_{\Sigma}^{(N)} = 30$~MeV, and
$U_{\Xi}^{(N)} =18$~MeV following Ref.~\cite{SG00}.

Inclusion of heavy $\sigma^*$ scalar meson introduces additional coupling constants and, thus,
hyperon potentials in the bath of hyperons are required to fit their values.
For this purpose, we follow the prescription suggested in Ref.~\cite{GHK14}. 
In hyperon matter composed of equal number of $\Xi^0$ and $\Xi^{-}$ only, the potential felt 
by the hyperons $Y$ at the saturation density can be written as
\begin{eqnarray}\label{eq:ydepth}
U_Y^{(\Xi)}  &=& g_{\omega Y}^{} \braket{\omega} +  g_{\phi Y}^{} \braket{\phi}
+ M_Y^{*} - M_Y 
\nonumber \\
&=& x_{\omega Y}^{} W + x_{\phi Y}^{} P - x_{\sigma Y}^{} S - x_{\sigma^* Y}^{} S^* .
\end{eqnarray}
Here, $S$, $W$, and $P$ can be obtained by solving Eqs.~\eqref{eq:ysig}, \eqref{eq:yome}, and 
\eqref{eq:yphi}. 
Then the combination of Eqs. \eqref{eq:ysigst} and \eqref{eq:ydepth} allows us to find the value of
$S^*$ and the corresponding $x_{\sigma Y}$. 
In the present work, we use the potential depths of hyperons as
$ U_{\Xi}^{(\Xi)} = U_{\Lambda}^{(\Xi)} = 2\,U_{\Xi}^{(\Lambda)} = 2\,U_{\Lambda}^{(\Lambda)} = -10 \,\text{MeV} $
following Refs.~\cite{SDGGS93,SDGGS94,GHK14}.

\section{Mass and radius of neutron stars}
\label{sec:nshyper}

The ground state of nuclear matter can be found by minimizing the energy density with respect to the
number density of its constituents. 
This gives the beta equilibrium conditions which lead to the relations of chemical potentials of particles as
\begin{equation}
\mu_i = \mu_n - q_i^{} \mu_e , \quad \mu_e = \mu_\mu,
\end{equation}
where $i$ represents $n$, $p$, $\Lambda^0$, $\Sigma^{\pm,0},$ and $\Xi^{0,-}$, 
while $q_i^{}$ stands for the charge of the baryon $i$.
The conservation of total baryon number density and charge neutrality lead to
\begin{eqnarray}
n_{b}^{} - \sum_{i} n_{i}^{} &=& 0 \,,\\
\sum_{i} q_{i}^{} n_i^{} - n_e^{} - n_\mu^{} &=& 0\,.
\end{eqnarray}

\begin{figure}[t]
\includegraphics[scale=0.45]{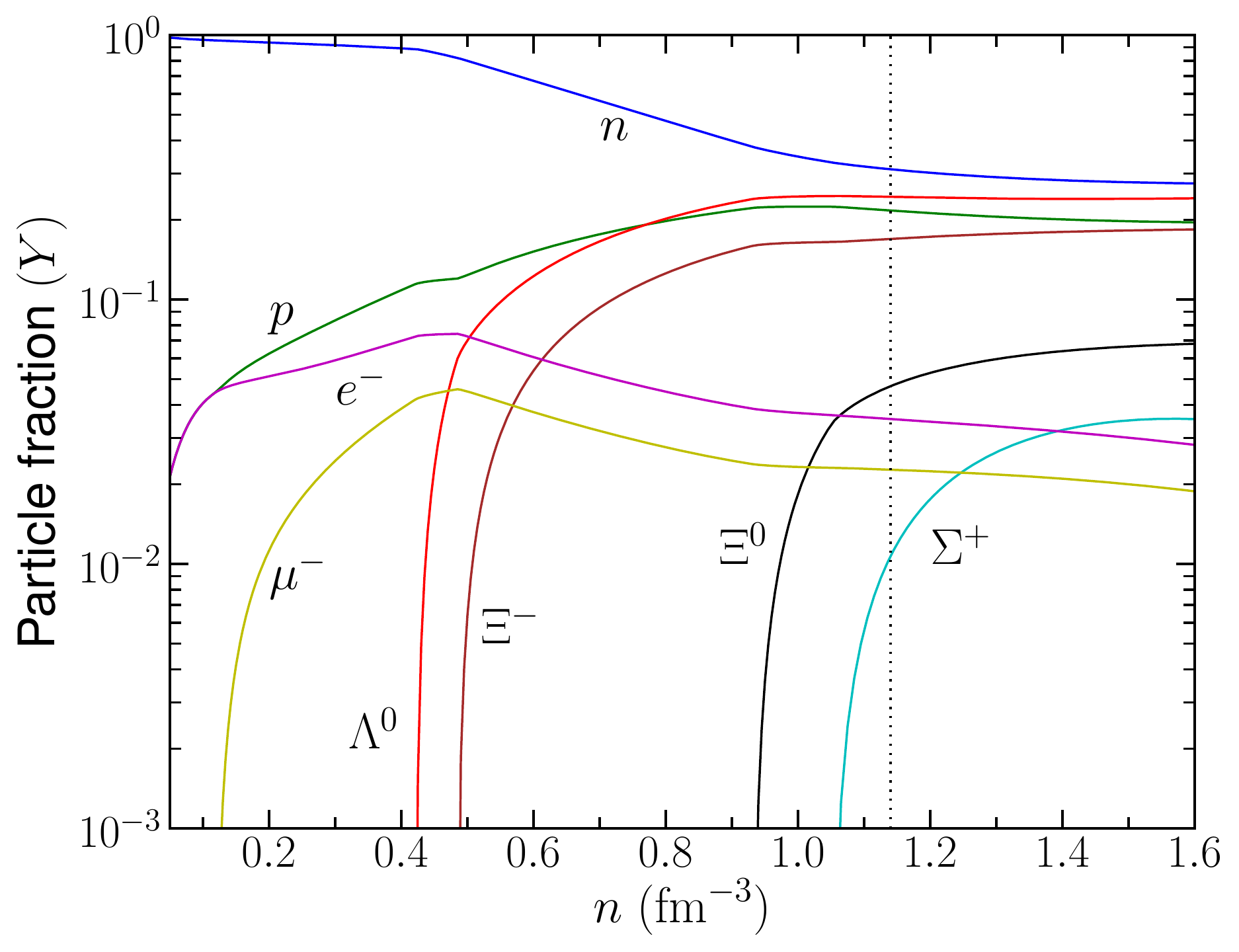}
\caption{
Particle fractions for given baryon number densities with $\mbox{RGCR} + \mbox{SU(6)}$ model, i.e., case I.}
\label{fig:gcr5_frac}
\end{figure}

\begin{figure}[t]
\includegraphics[scale=0.45]{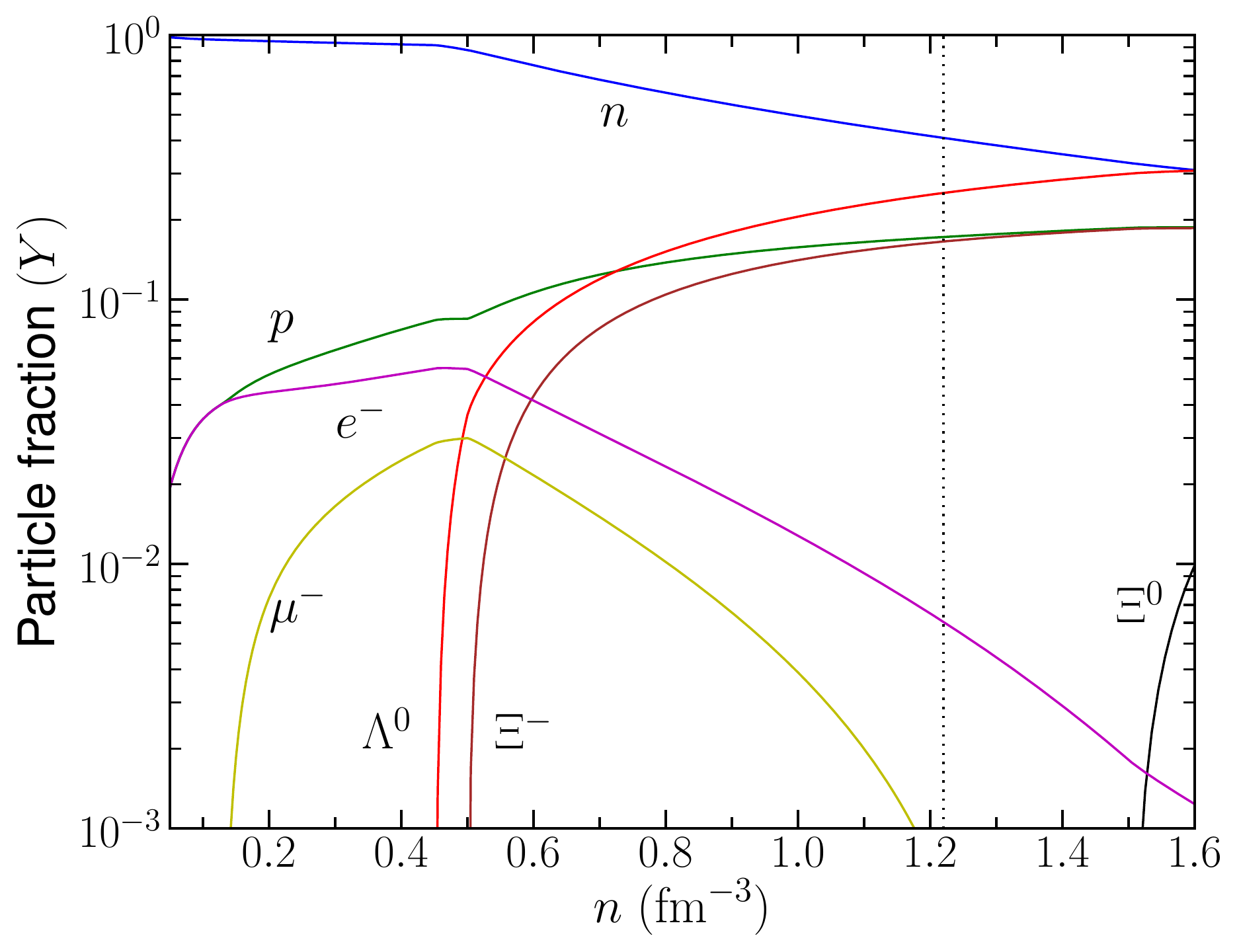}
\caption{
Particle fractions for given baryon number densities with 
$\mbox{RDSS} + \mbox{SU(6)}$ model, i.e., case I.}
\label{fig:ds2_frac}
\end{figure}

Shown in Figs.~\ref{fig:gcr5_frac} and \ref{fig:ds2_frac} are the particle fractions 
in the beta-stable nuclear matter. 
The vertical dotted line in each graph indicates the central baryon number density 
in the maximum mass of a neutron star in each model.
In the $\mbox{RDSS}+\mbox{SU(6)}$ model, there can exist $\Lambda^0$ and
 $\Xi^-$ in the core of neutron stars. 
On the other hand, it is possible to have $\Lambda^{0}$, $\Xi^-$, $\Xi^{0}$, and 
$\Sigma^+$ in the core of neutron stars if we use RGCR model with case I.

\begin{figure}[t]
\includegraphics[scale=0.45]{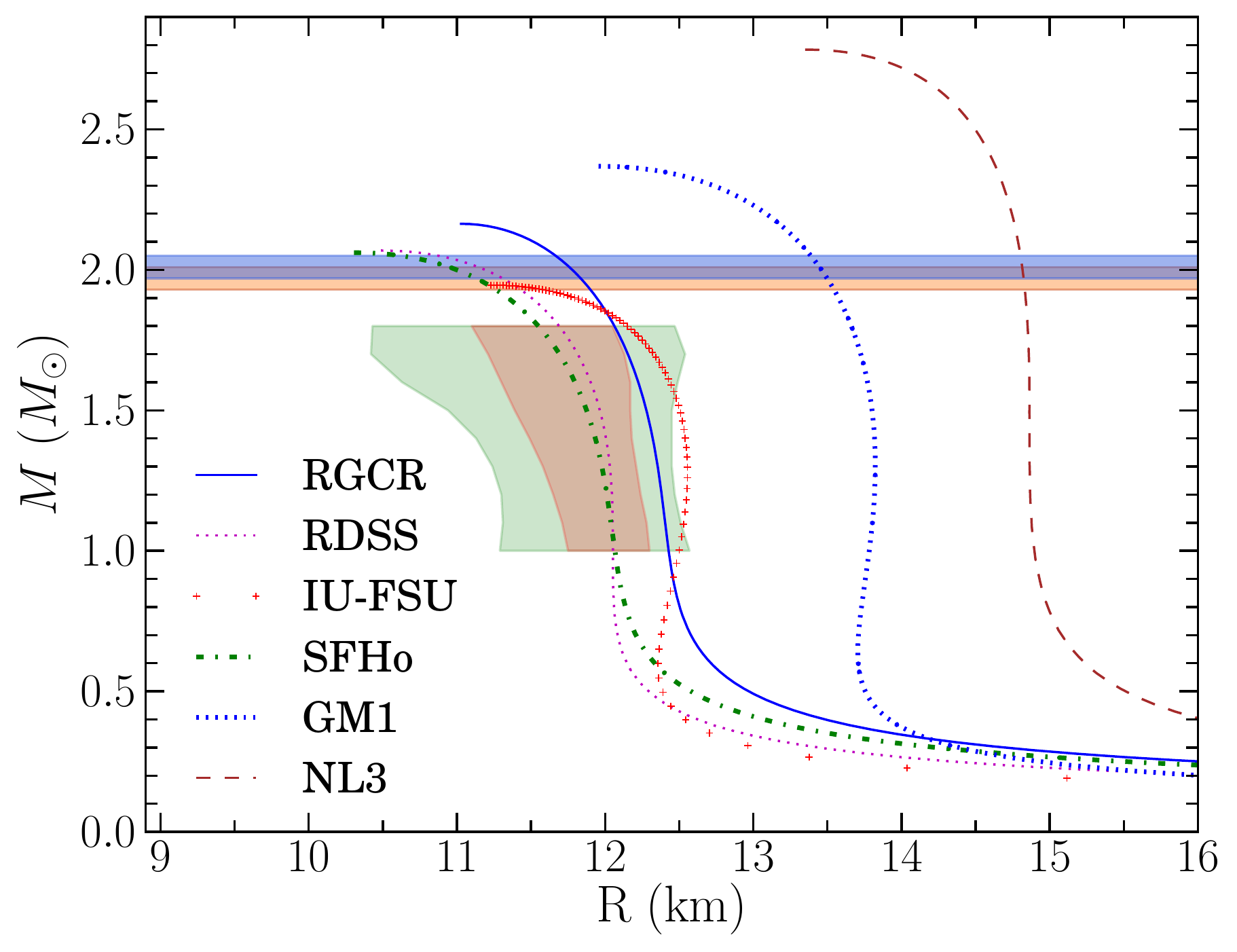}
\caption{
Mass and radius of neutron stars using relativistic mean field models
without hyperons. 
RGCR and RDSS models are the results of our
calculation and the other models are explained in Table~\ref{tb:rmf_sym}.
The horizontal lines indicate the observed neutron star masses of Ref.~\cite{DPRRH10, AFWT13}.
The brown and green shaded areas show the allowed region of the mass-radius constraint of 
Ref.~\cite{SLB10} at the $1 \sigma$ and $2 \sigma$ level, respectively.
}
\label{fig:rmf_nsmr}
\end{figure}

For the crust EOS, we use the liquid drop model approaches as explained in Ref.~\cite{LH17}
using the SLy4 force model~\cite{CBHMS98}.
For a given EOS, the mass and radius relation of neutron stars is obtained by solving the
Tolman-Oppenheimer-Volkoff equation,
\begin{eqnarray}
\frac{dP(r)}{dr} &=& -\frac{G m(r)}{r^2}
\left[\mathcal{E}(r) + \frac{P(r)}{c^2}\right]
\left[1 + \frac{4\pi r^3 P(r) }{m(r) c^2}\right] 
\nonumber \\ && \mbox{} \times 
\left[1 -\frac{2Gm(r)}{rc^2}\right]^{-1}
\end{eqnarray}
with
\begin{eqnarray}
\frac{d m(r)}{dr} &= & 4\pi \mathcal{E}(r)\,r^2 .
\end{eqnarray}
Figure~\ref{fig:rmf_nsmr} shows the obtained mass-radius relations using the relativistic mean field 
models discussed in the present work. 
In this figure, the horizontal lines indicate the observed neutron star masses of 
Ref.~\cite{DPRRH10, AFWT13}.
The brown and green areas show the empirical region of the mass-radius constraint given in
Ref.~\cite{SLB10} with the $1 \sigma$ and $2 \sigma$ level, respectively.
This figure shows that all the considered models in the present work can satisfy the criterion given by
the neutron star mass.
However, the GM1 and NL3 models are found to yield very large neutron star radii 
compared with the empirically allowed region of Ref.~\cite{SLB10}.
This is because  these models have large nuclear incompressibility ($K$) and, in particular, large density gradient ($L$)
of the nuclear symmetry energy.
These results emphasize the important role of the combined mass-radius constraint to understand the
EOS of neutron stars.

We now discuss our results based on the SU(3) models that show the role of strangeness in
the structure of a neutron star.
We first examine the maximum neutron star mass allowed by each model and the obtained results are 
shown in Table~\ref{tb:hystar}.
We present the results for given values of $z$ and $\alpha_V^{}$ for each model. 
In general, the existence of hyperons reduces the maximum mass of neutron stars, which confirms the
phenomenon known as the hyperon puzzle.
In particular, the reduction of the maximum neutron star mass is large when we use the SU(6) relations
for couplings, i.e., the case I, as shown by the second row of Table~\ref{tb:hystar}.
Even in this case, the GM1 and NL3 models give large values for neutron star mass.
However, as mentioned before, these models result in neutron star radius that is much larger than the
empirically allowed values.
We then vary the values of $z$ and $\alpha_V^{}$ assuming the SU(3) symmetry relations for
coupling constants.
In this case, since the vector meson ($\omega$) self interaction exists in the IU-FSU and SFHo models,
we assume $g_{\phi N}=0$ to use the rescaling equation given in Eq.~\eqref{eq:rescale}.
Shown in the third row of Table~\ref{tb:hystar} are the results of case II with a reduced $z$ value compared with the
SU(6) case.
The fourth row of Table~\ref{tb:hystar} shows the results of case III by with $\alpha_V^{} = \frac12$.
Compared with the SU(6) models (case I), it is evident that the models with SU(3) symmetry (cases II and III)
are less constrained by the group structure and the degree of the hyperon puzzle is reduced very much.
In fact, the models of RGCR and RDSS can meet the mass condition of neutron stars when we
use the SU(3) relations and varying the values of $\alpha_V^{}$ and $z$.
In principle, one cannot simply apply the SU(3) relations in case of
IU-FSU and SFHo models because of non-vanishing $\zeta$.
Since the values in Table~\ref{tb:hystar} are obtained by applying the SU(3) relations, they are given
in parentheses.
Therefore, the maximum masses of neutron stars in case of II and III are subject to change by more
realistic calculations.

\begin{table}[t]
\caption{The maximum mass of neutron stars (in units of $M_{\odot}$) in each model
using $U_{\Lambda}^{(N)} = -30~\si{MeV}$, $U_{\Sigma}^{(N)} = +30~\si{MeV}$, 
$U_{\Xi}^{(N)} = -18~\si{MeV}$ and $U_{\Xi}^{(\Xi)} = U_{\Lambda}^{(\Xi)} 
= 2U_{\Xi}^{(\Lambda)} = 2U_{\Lambda}^{(\Lambda)} = -10~\si{MeV}$.
Note that SFHo and IU-FSU have non-vanishing $\zeta$ thus 
the maximum mass of neutron stars in case of II and III is not physical.
}
\begin{center}
\begin{tabular}{ccccccccc}
\toprule
Model        & $z$           & $\alpha_V^{}$   & RGCR          & RDSS & IU-FSU     & SFHo  &  GM1   & NL3 \\
\hline 
SU(2)         & ---         & ---            
					  & 2.22  
                                            & 2.07  
                                            & 1.94  
                                            & 2.06  
                                            & 2.36  
                                            &  2.78  
\\[1mm]
Case I & $\frac{1}{\sqrt{6}}$     & $1$            
					  & 1.78 
                                            & 1.71 
                                            &  1.67 
                                            &  1.70
                                            & 1.93 
                                            &  2.25 
\\[1mm]
Case II & $\frac{1}{2\sqrt{6}}$  & $1$            
					  & 2.03 
                                            & 1.90 
                                            & (1.93) 
                                            & (1.88) 
                                            & 2.15 
                                            &  2.26  
\\[1mm]
Case III & $\frac{1}{\sqrt{6}}$      & $\frac12$          
					  & 1.98 
                                            & 1.91 
                                            & (2.03) 
                                            & (1.88) 
                                            & 2.14 
                                            &  2.51 
\\[1mm]
\hline
\end{tabular}
\end{center}
\label{tb:hystar}
\end{table}

\begin{figure}[t]
\includegraphics[scale=0.45]{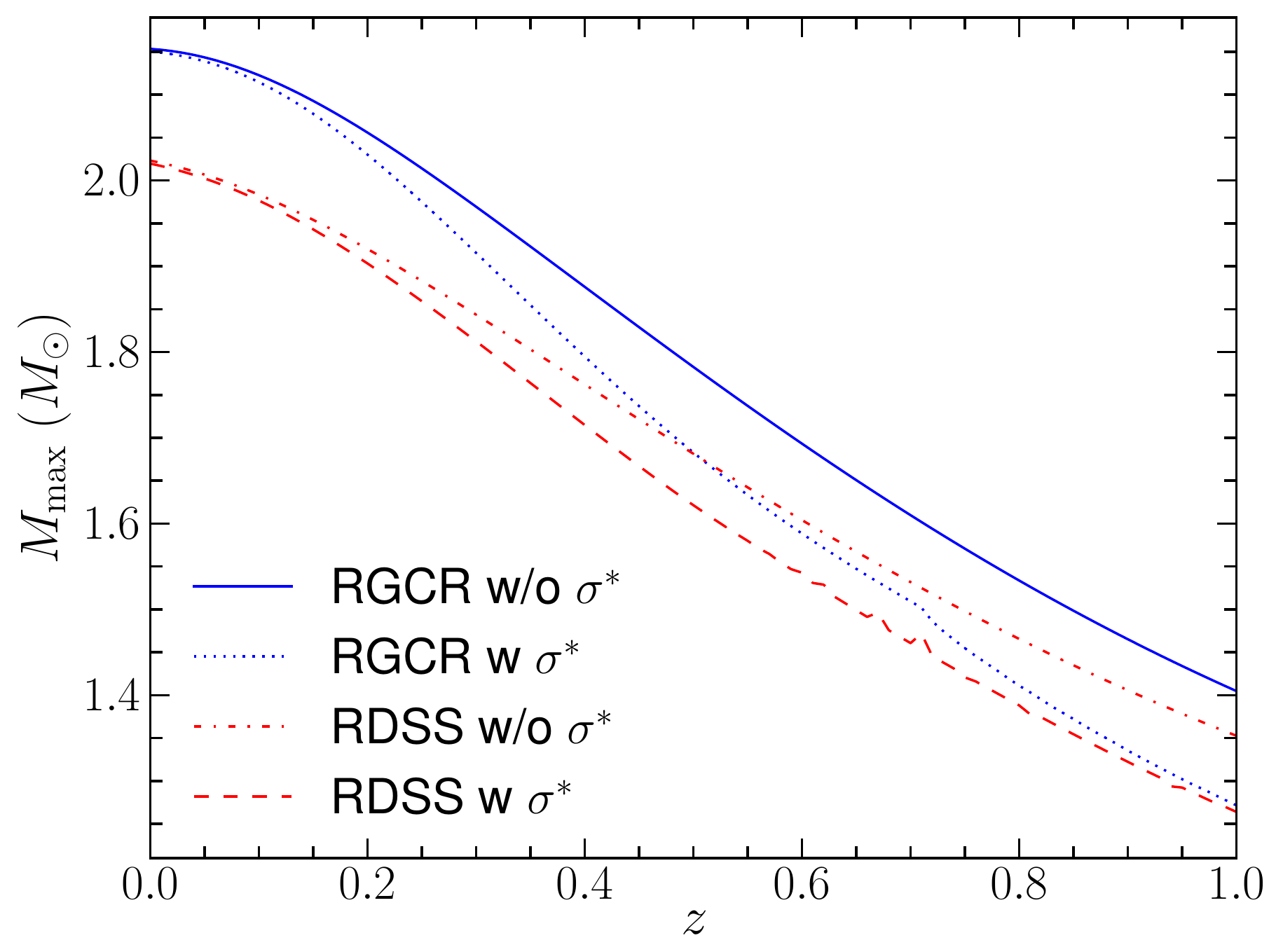}
\caption{
Maximum mass of neutron stars as a function of $z$ in case II with $\alpha_V^{}=1$.}
\label{fig:zmass}
\end{figure}

\begin{figure}[t]
\includegraphics[scale=0.45]{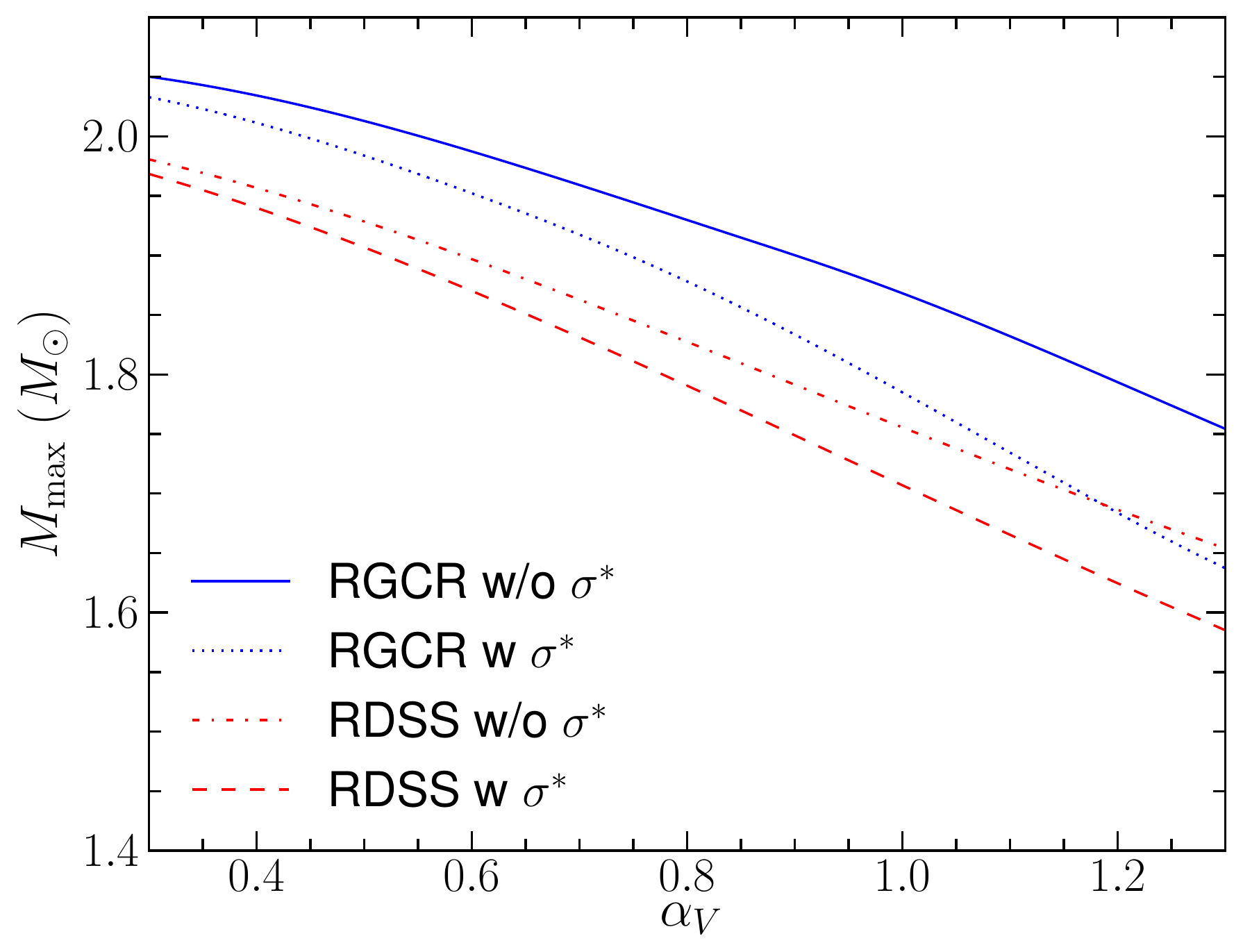}
\caption{
Maximum mass of neutron stars as a function of $\alpha_V^{}$ in case III  with $z=1/\sqrt{6}$.}
\label{fig:amass}
\end{figure}

More detailed results on the dependence of the maximum neutron star mass on the couplings are shown
in Figs.~\ref{fig:zmass} and \ref{fig:amass}.
The results of case II are presented in Fig.~\ref{fig:zmass} with varying the value of $z$ from $0$ to $1$.
Those of case III are shown in Fig.~\ref{fig:amass} with $0.3 \le \alpha_V^{} \le 1.3$.
We find that the maximum mass of neutron stars decreases as $z$ or $\alpha_V^{}$ increases in case II 
and case III, respectively.
This observation confirms the results of Ref.~\cite{WCS12b}.
To achieve $2 M_\odot$ for the neutron star mass, we need $z \le 0.3$ in case II and $\alpha_V^{} \le 0.5$ in case III for
the RGCR model.
The RDSS model requires even smaller values for $z$ and $\alpha_V^{}$.
In Figs.~\ref{fig:zmass} and \ref{fig:amass}, we also show the results with and without $\sigma^*$ to
find that the presence of the $\sigma^*$ reduce further the maximum mass of neutron stars.

\begin{figure}[t]
\includegraphics[scale=0.45]{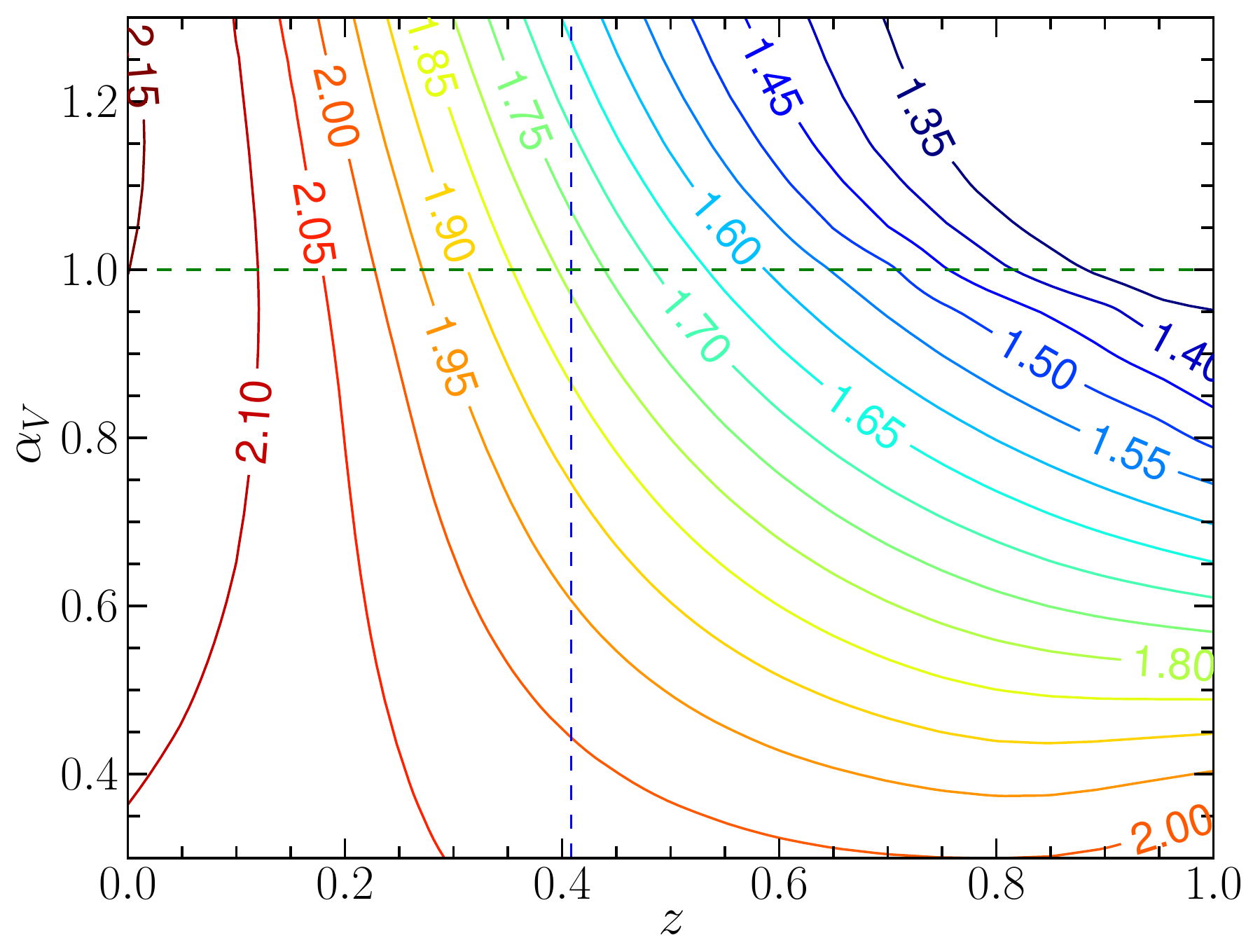}
\caption{
Contour plot of the maximum mass of neutron stars as a function of
$z$ and $\alpha_V^{}$ in the RGCR model with case IV.
The horizontal and vertical dashed lines are $\alpha_V^{} = 1.0$ and $z = 1/\sqrt6$, respectively.
}
\label{fig:gcr5_cont}
\end{figure}

\begin{figure}[t]
\includegraphics[scale=0.45]{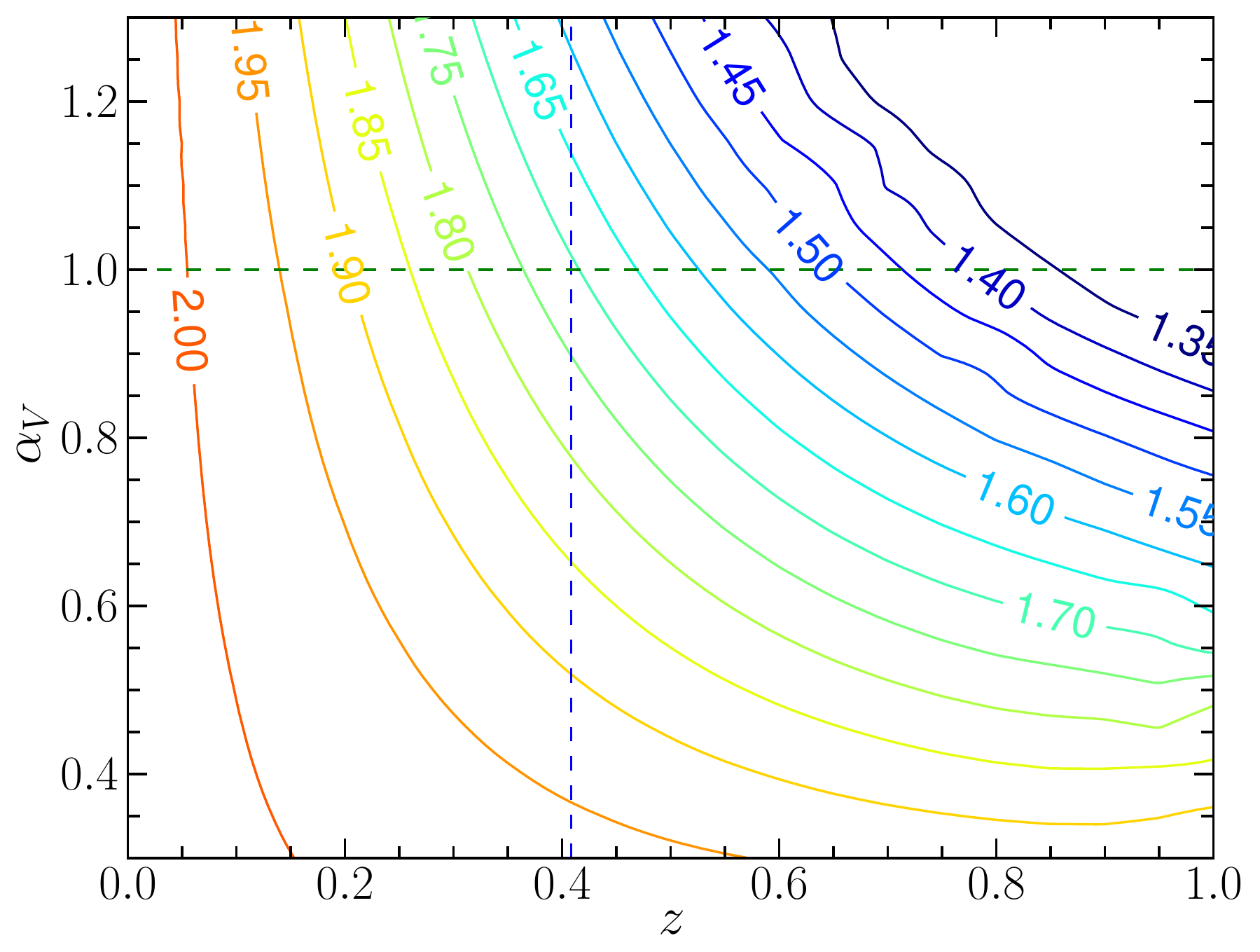}
\caption{
Contour plot of the maximum mass of neutron stars as a function of
$z$ and $\alpha_V^{}$ in the RDSS model with case IV.
The horizontal and vertical dashed lines are $\alpha_V^{} = 1.0$ and $z = 1/\sqrt6$, respectively.
}
\label{fig:dss2_cont}
\end{figure}

For case IV, where we vary both $z$ and $\alpha_V$, the results are presented as a contour plot 
in Fig.~\ref{fig:gcr5_cont} for the RGCR model and in Fig.~\ref{fig:dss2_cont} for the RDSS model . 
The horizontal dashed lines represent $\alpha_V = 1$ and correspond to case II, while the vertical
dashed lines denote $z = 1/\sqrt6$ corresponding to case III.
As expected from the results shown in Figs.~\ref{fig:zmass} and \ref{fig:amass}, small values for 
$z$ and $\alpha_V^{}$ are needed to allow for massive neutron stars.
Our results show that if the $\alpha_V^{}$ ratio of the vector meson couplings is the same 
as in the free space, the coupling ratio between octet vector meson and singlet vector meson 
should change from $1/\sqrt6 \approx 0.4$ to about $0.3$.
On the other hand, if the coupling ratio is kept as $1/\sqrt6$, the value of $\alpha_V^{}$ should 
be reduced to below $0.45$.
It is interesting to note that this value is close to the $\alpha$ value of pseudoscalar mesons, 
of which free space value is estimated to be $\alpha_{\rm PS}^{} = 0.355$ in Ref.~\cite{RSY99}.
More rigorous investigations on the change of couplings in dense nuclear matter are, therefore,
highly desirable, in particular, for both $\alpha_V^{}$ and $\alpha_{\rm PS}^{}$.

\begin{figure}[t]
\includegraphics[scale=0.45]{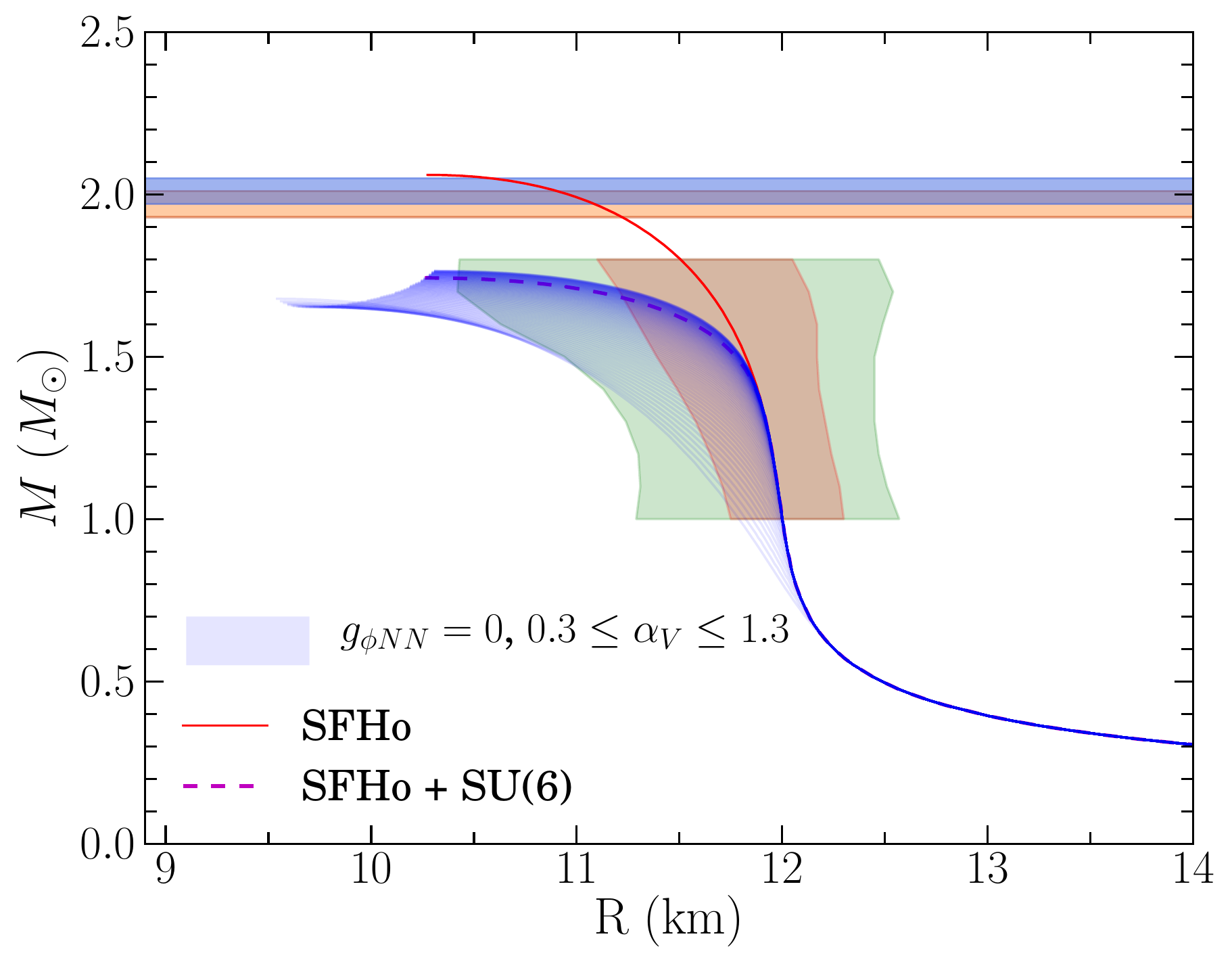}
\caption{
Mass and radii curves with the variation of $\alpha_V^{}$ constrained by $g_{\phi N}^{} = 0$ in 
the SFHo model with case I. Smaller $\alpha_V^{}$ gives smaller maximum mass of a neutron star for given EOS.}
\label{fig:sfho_hyp}
\end{figure}

\begin{figure}[t]
\includegraphics[scale=0.45]{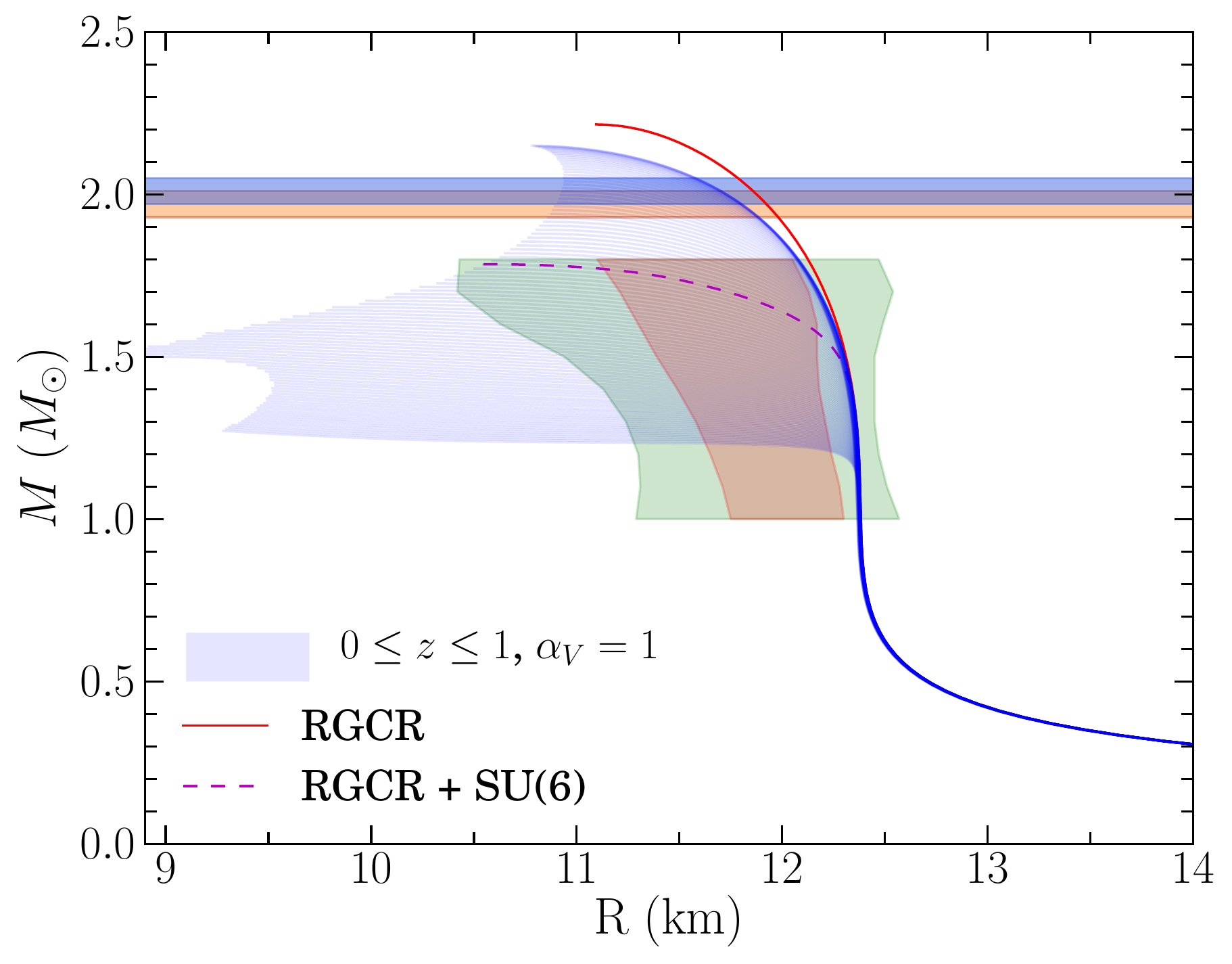}
\caption{
Mass and radii curves with the variation of $z$ with $\alpha_V^{} =1$ in the RGCR model, i.e., case II.
The red solid line is the result of the model in the SU(2) case and the dashed line is that of case I.
}
\label{fig:gcr5su3z}
\end{figure}

\begin{figure}[t]
\includegraphics[scale=0.45]{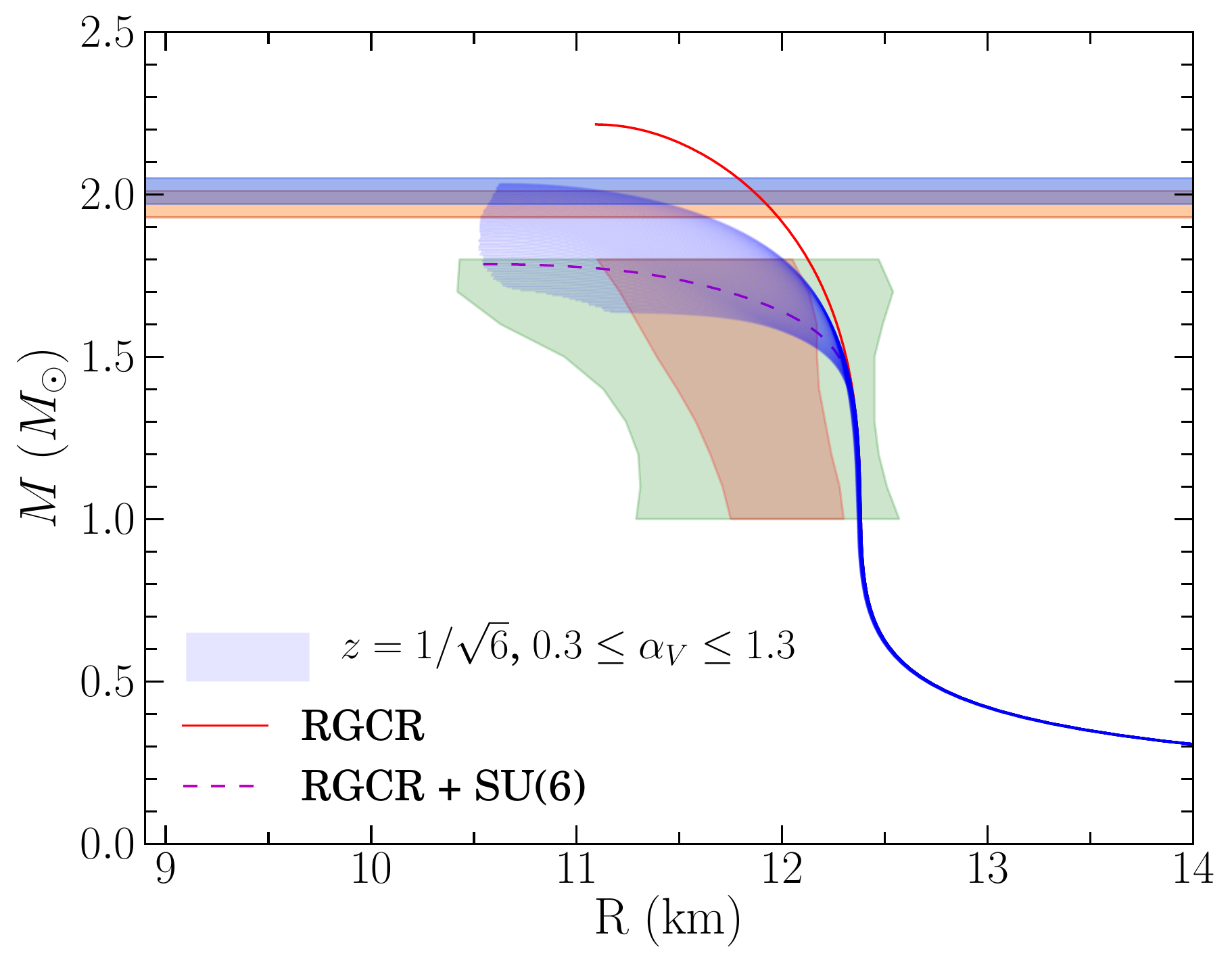}
\caption{
Same as Fig.~\ref{fig:gcr5su3z} but with the variation of $\alpha_V$ with $z=1/\sqrt{6}$ in the RGCR model, i.e., case III.
}
\label{fig:gcr5su3a}
\end{figure}

\begin{figure}[t]
\includegraphics[scale=0.45]{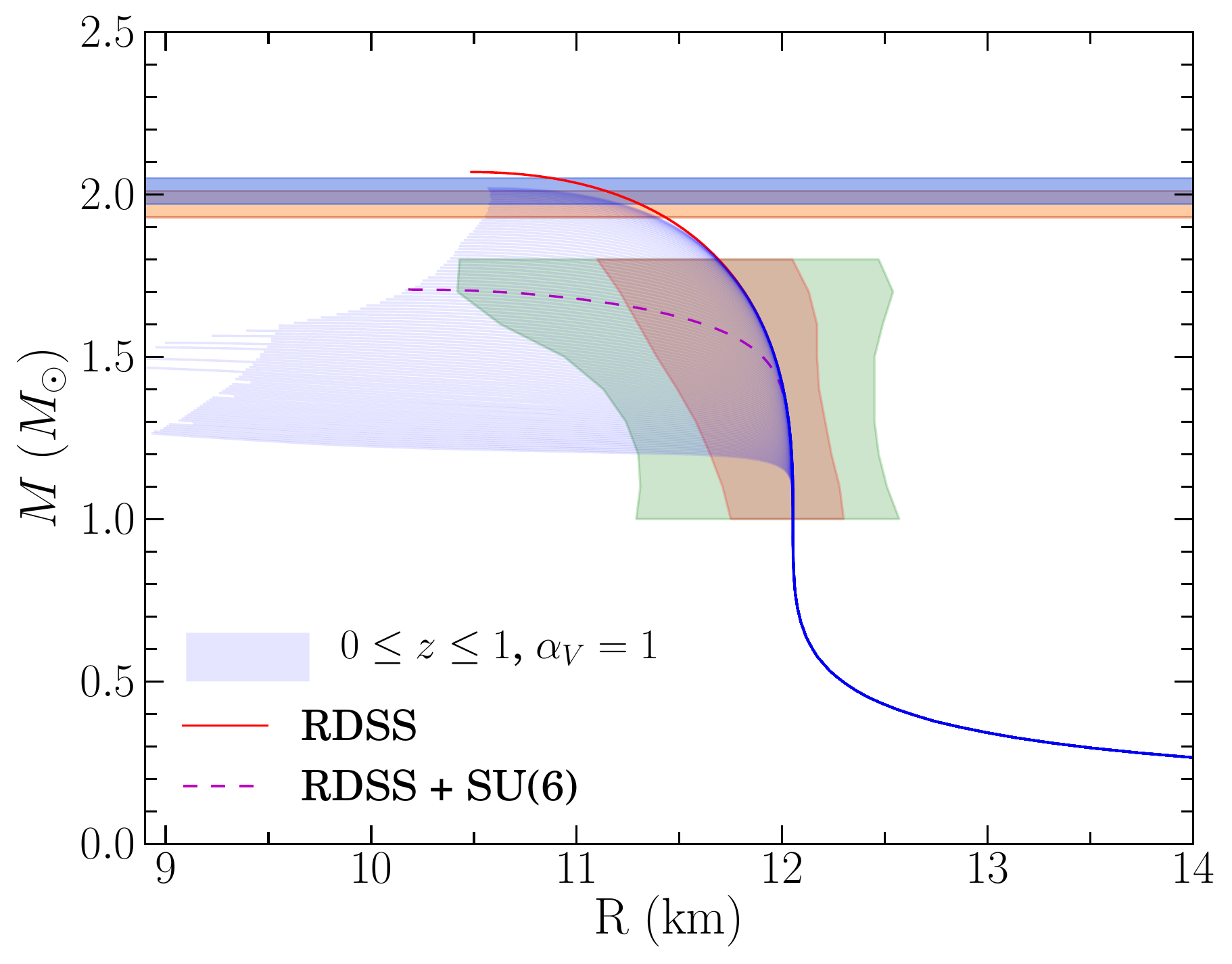}
\caption{
Same as Fig.~\ref{fig:gcr5su3z} but in the RDSS model.
}
\label{fig:dss2su3z}
\end{figure}

\begin{figure}[t]
\includegraphics[scale=0.45]{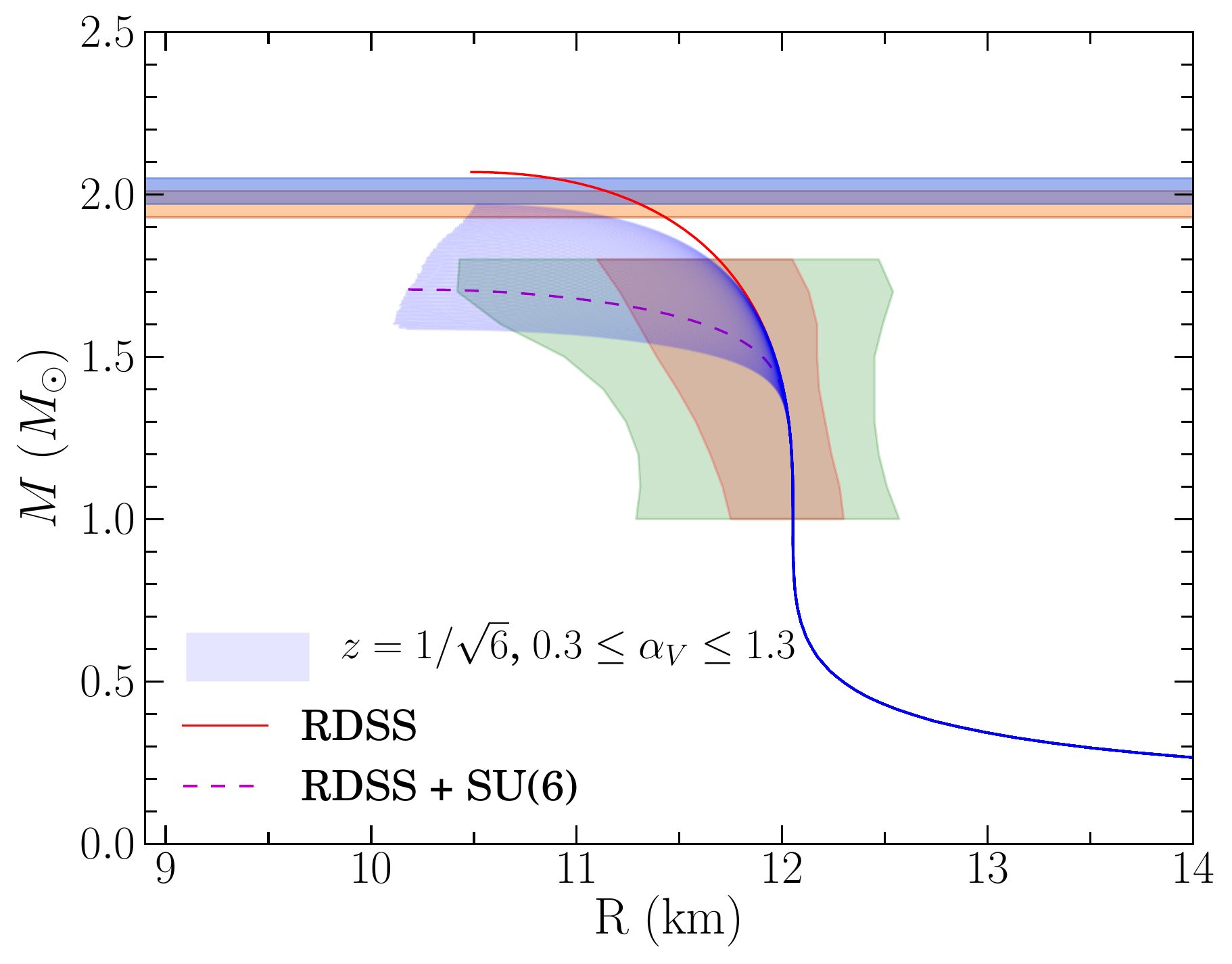}
\caption{
Same as Fig.~\ref{fig:gcr5su3a} but in the RDSS model.
}
\label{fig:dss2su3a}
\end{figure}

Since the purpose of this article is to see the role of the mass-radius constraint for neutron star models,
we now explore the model-dependence of the predicted mass-radius region of neutron stars of each model. 
We vary either $\alpha_V^{}$ or $z$ and denote the range of the obtained mass-radius curves by blue shaded
areas in Figs.~\ref{fig:sfho_hyp}--\ref{fig:dss2su3a}.
Since, as shown in Fig.~\ref{fig:rmf_nsmr}, the GM1 and NL3 models in the SU(2) case cannot satisfy the
empirically allowed mass-radius region and the maximum neutron star mass of the IU-FSU model is  
smaller than $2 M_\odot$, we focus on the SFHo, RGCR, RDSS models in the followings.

Figure~\ref{fig:sfho_hyp} shows the mass and radius curves from the SFHo RMF 
model with hyperons. 
Its parameters in the SU(2) sector are given in Table~\ref{tb:rmfpa}. 
The blue region is obtained with the variation of $\alpha_V$ constrained by $g_{\phi N}^{} = 0$. 
This shows that, although the SFHo model can pass the criterion for the radius, the predicted 
maximum mass cannot achieve 2.0~$M_\odot$.
Since the $\phi$ meson gives a repulsion between baryons, the condition that $g_{\phi N}^{} \ne 0$ 
may give a larger mass for neutron stars.
However, since the SFHo model barely satisfies 2.0~$M_\odot$ condition for the neutron 
star mass within the SU(2) configuration, the mass reduction from the hyperons does not allow the model
to fulfill the maximum mass criteria in any SU(3) models.

Shown in Fig.~\ref{fig:gcr5su3z} are the area of mass-radius of neutron stars obtained in the RGCR 
model by varying the value of $z$ as $0 \le z \le 1$ while keeping $\alpha_V^{} = 1$, which corresponds
to case II.
On the other hand, Fig.~\ref{fig:gcr5su3a} shows the results with $0.3 < \alpha_V < 1.3$ while keeping $z=1/\sqrt{6}$.
Therefore, it corresponds to case III. 
The results presented in Figs.~\ref{fig:dss2su3z} and \ref{fig:dss2su3a} are obtained with the RDSS model
for case II and case III, respectively.
Because $g_{\phi N} \ne 0$ and the maximum mass is greater than 2.0~$M_\odot$ in the SU(2) case for the both models, 
the predicted mass-radius curves have a chance to fulfill the empirical constraints on the mass and radius of neutron stars.
These results indicate that it would be possible to have hyperons in the core of neutron stars by satisfying
the maximum mass criteria when a proper nucleon EOS, i.e., in the SU(2) model, is used.
They also imply that the understanding of the changes of $\alpha_V^{}$ and $z$ couplings in nuclear matter and 
hyperon matter would shed light on resolving the hyperon puzzle.
Our results also show that the case III, Figs.~\ref{fig:gcr5su3a} and \ref{fig:dss2su3a}, predicts a narrower area of 
mass-radius curves compared with case II, Figs.~\ref{fig:gcr5su3z} and \ref{fig:dss2su3z}.
This means that the $z$-dependence of the mass-radius curve is more sensitive than its $\alpha_V^{}$ dependence.

\section{Hyperon direct Urca process}
\label{sec:urca}

The hyperon direct Urca process plays an important role in the thermal history of a 
neutron star~\cite{HG94b, SBS98,LHL16}
because the appearance of hyperons allows the hyperon direct Urca process
at a relatively small proton fraction~\cite{PPLP92}.
Various types of hyperon direct Urca are allowed according to the number of hyperon species~\cite{PPLP92}.
The EOS with small $L$ ($<45$ MeV) may not allow nucleon direct Urca process even with 
the maximum mass of neutron stars~\cite{LHL15} because the proton fraction does not 
increase fast enough to turn on the direct Urca process as the baryon number density increases
in the core of neutron stars. 
However, inclusion of hyperons might change the situation for direct Urca process both for
nucleons and hyperons through
\begin{equation}
\begin{aligned}
B_1 \rightarrow B_2 + l + \bar{\nu}_l^{} \\
B_2 + l \rightarrow B_1 + \nu_l^{} ,
\end{aligned}
\end{equation}
where $B_1$ and $B_2$ are baryons and $l$ denotes a lepton and $\nu_l$ is the neutrino associated with
the lepton $l$.

Table~\ref{tb:urca_sfho} shows the critical baryon number density 
and mass of neutron stars for the hyperon direct Urca process
in the case of the SFHo model in case I. 
Table~\ref{tb:rgcr} presents the same quantities but in the RGCR model in case II.
This indicates that the hyperon direct Urca process turns on earlier than the nucleon
direct Urca process.
In general, a small amount of hyperons, compared with the proton fraction for the nucleon direct Urca
process, is able to turn on the hyperon direct Urca processes.
For example, in the case of the SFHo model, the nucleon direct Urca process does not occur
even in the maximum mass (2.06~$M_\odot$) of the neutron star, 
while the hyperon direct Urca process turns on when the
neutron star mass is larger than $1.26~M_\odot$ as shown in Table~\ref{tb:urca_sfho}.

\begin{table}[t]
\caption{The critical density and critical mass of neutron stars for
the baryon direct Urca process in the SFHo model in case I.}
\label{tb:urca_sfho}
\begin{tabular}{ccc}
\hline\hline
Urca  & $n_c^{} (\si{fm}^{-3})$  & $M_c (M_\odot)$ \\
\hline
$\Lambda \rightarrow p + e + \bar{\nu}_e $            & 0.470 & 1.26 \\
$\Lambda \rightarrow p + \mu + \bar{\nu}_\mu $        & 0.475 & 1.28 \\
$\Xi^{-} \rightarrow \Lambda + e + \bar{\nu}_e $      & 0.516 & 1.38 \\
$\Xi^{-} \rightarrow \Lambda + \mu + \bar{\nu}_\mu $  & 0.515 & 1.38 \\
\hline
\end{tabular}
\end{table}

The maximum mass of neutron star which contains hyperons in case of SFHo model, 
however, is much less than $2.0\,M_\odot$ as shown in Table~\ref{tb:hystar}.
This may indicate that strong repulsion between nucleons or many-body forces
in the nuclear EOS is preferred to make high enough maximum 
mass of neutron stars so that the mass reduction resulting from the existence of 
hyperons is consistent with the observed maximum mass.
For instance, in the RGCR model with $z=1/(2\sqrt{6})$ and $\alpha_V^{}=1$,
the maximum neutron star mass is predicted to be $2.03~M_\odot$ as shown in Table~\ref{tb:hystar}.
The first hyperon direct Urca process happens when $M=1.47~M_{\odot}$,
and it allows the nucleon direct Urca process when $M=1.58~M_{\odot}$.
In this model, $\Lambda$ hyperons appear if the baryon number density is greater
than $n > 0.468 ~\mathrm{fm}^{-3}$ and the direct Urca process involving
$\Lambda$ occurs if $n > 0.470 ~\mathrm{fm}^{-3}$, which corresponds to 
$\rho_\Lambda^{} = 2.62\times 10^{-4} ~\mathrm{fm}^{-3}$ that is
$0.06\%$ to the total baryon number density.
As shown in these examples, if hyperons exist in the core of neutron stars,
the hyperon direct Urca process may happen at smaller baryon number density than the density required
for the nucleon direct Urca process.

\begin{table}[t]
\caption{The critical density and critical mass of neutron stars for
the baryon direct Urca process in case of the RGCR model in case II with $z=1/(2\sqrt{6})$ and $\alpha_V^{}=1$.
}
\begin{tabular}{ccc}
\hline\hline
Urca  & ~$n_c$ ($\mathrm{fm}^{-3}$)~  & ~$M_c$ ($M_\odot$)~ \\
\hline
$\Lambda \rightarrow p + e + \bar{\nu}_e $            & 0.470 & 1.47 \\
$\Lambda \rightarrow p + \mu + \bar{\nu}_\mu $        & 0.475 & 1.49 \\
$n \rightarrow p + e + \bar{\nu}_e$                   & 0.507 & 1.58\\
$\Xi^{-} \rightarrow \Lambda + e + \bar{\nu}_e $      & 0.555 & 1.70 \\
$\Xi^{-} \rightarrow \Lambda + \mu + \bar{\nu}_\mu $  & 0.555 & 1.70 \\
$n \rightarrow p + \mu + \bar{\nu}_\mu$               & 0.589 & 1.76\\
$\Xi^{-} \rightarrow \Xi^0 + e + \bar{\nu}_e $        & 0.949 & 2.04 \\
$\Xi^{-} \rightarrow \Xi^0 + \mu + \bar{\nu}_\mu $    & 0.965 & 2.05 \\
$\Xi^{0} \rightarrow \Sigma^0 + e + \bar{\nu}_e $     & 0.987 & 2.05 \\
$\Xi^{0} \rightarrow \Sigma^0 + \mu + \bar{\nu}_\mu $ & 0.976 & 2.05 \\
\hline
\end{tabular}\label{tb:rgcr}
\end{table}

\section{Conclusion}
\label{sec:con}

In the present work, we investigated the role of hyperons in neutron stars within 
the relativistic mean field approach.
For this purpose, we first constructed SU(2) relativistic mean field models, RGCR and RDSS, 
whose parameters are determined by symmetric nuclear matter properties and theoretical 
calculations for pure neutron matter.
We found that these models satisfy the mass-radius constraints of neutron stars of
Ref.~\cite{SLB10}. 
The extension to the flavor SU(3) was then made to account for the contribution of hyperons 
in the energy density and pressure of baryon matter.
In general, the existence of hyperons makes the EOS softer than those with nucleons only,
which makes the neutron star containing hyperons always less massive than that without hyperons.
To understand the reduction of neutron star mass due to hyperons, we analyze the RMF models 
in the literature as well as those developed in the present work.
The potential depths of hyperons ($\Lambda$, $\Sigma$, $\Xi$) at the saturation nucleon density
are used to obtain the hyperon coupling constants. 
We then tested the effects of the variation of couplings, $\alpha_V^{}$ and $z$, to neutron star's masses
and radii.
We found that the RGCR and RDSS models can satisfy the mass-radius constraints of neutron stars
with certain values of $\alpha_V^{}$ and $z$.
However, other models in the literature have difficulties to fullfill the mass-radius constraints even with
the variation of the couplings. 
This shows that it is needed to take into account the pure neutron matter properties for determination of
the SU(2) parameters in order to explain the observed neutron star properties.
Furthermore, rigorous investigations on the change of SU(3) coupling constant parameters in dense matter
are required to understand the structure of neutron stars in depth.

The presence of hyperons is supposed to change the cooling history of neutron stars since 
the condition for hyperon direct Urca process is not restrictive as in the case of nucleon direct 
Urca process~\cite{PPLP92}. 
The hyperon direct Urca process affects the cooling of neutron stars which contain hyperons in the core.
Thus neutron stars whose masses are greater than the critical mass with the existence of hyperon 
should have different cooling history.
In the present work, we estimated the critical density and critical neutron star mass for hyperon
direct Urca processes.
Our results indicate that neutron star with a mass greater than $1.5\,M_\odot$ is likely to turn on the 
hyperon direct Urca process.

\acknowledgments
CHL was supported by the National Research Foundation (NRF) of Korea
grant funded by the Korea government (MISP) under Grant Nos.
NRF-2015R1A2A2A01004238 and NRF-2016R1A5A1013277.
Y.O. was supported by the NRF of Korea under Grant Nos. NRF-2015R1D1A1A01059603 
and NRF-2016K1A3A7A09005577.

\appendix
\section{Relativistic mean field model and nuclear properties}

As was discussed in Sec.~\ref{sec:nuc}, we determine the parameter values of the RMF model 
using the properties of symmetric nuclear matter given in Eq.~(\ref{eq:matter_property}), namely,
$M_N^*$, $B$, $n_0$, and $K$.
With the Lagrangian $\mathcal{L}_{\sigma\omega\rho}$ given in Eq.~\eqref{eq:lag},
these quantities are obtained as
\begin{align}
& M^*  = M - S_0\,,\\
& \mu_n^{}  = \mu_p^{} = \sqrt{k_F^2 + M^{*2}} + W_0\,, \\ 
& \begin{aligned}
B  = & M - \frac{1}{n_0}
 \biggl[ V(S_0) + \frac{1}{2}\left(\frac{m_\omega}{g_\omega}\right)^2W_0^2
 + \frac{\zeta}{8}W_0^4 \\
 & \qquad\qquad
 + \frac{2}{\pi^2}\int_{0}^{k_F}dk \,k^2 \sqrt{k^2 + M^{*2}}\biggr]\,, \label{eq:bind} 
 \end{aligned}
\\
& \begin{aligned}
P(n_0)  = & - V(S_0) +
\frac{1}{2}\left(\frac{m_\omega}{g_\omega}\right)^2W_0^2
+ \frac{\zeta}{24}W_0^4 \\
& + \frac{2}{3\pi^2}\int_{0}^{k_F}dk\, 
\frac{k^4}{\sqrt{k^2 + M^{*2}}} \,, \label{eq:pres}
\end{aligned}
\\
& K  =  9\, n_0\left[ \left(\frac{m_\omega}{g_\omega}\right)^2 
       + \frac{\zeta}{2}W_0^2\right]^{-1} 
    + 3 \frac{k_F^2}{E_F^*}  \label{eq:comp} \\
   & \quad
    - 9\, n_0\left(\frac{M^*}{E_F^*}\right)^2 
  \left[\left( \frac{\partial^2}{\partial S^2} + \frac{3}{M^*}\frac{\partial}{\partial S}
       \right)_{0}V(S) -3\frac{n_0}{E_F}\right]^{-1}\,, \notag
\end{align}
at the saturation density, where
\begin{equation}
V(S) = \frac{1}{2}\left( \frac{m_\sigma^{}}{g_\sigma^{}} \right)^2 S^2
+ \frac{\kappa}{3!}S^3 + \frac{\lambda}{4!}S^4\,.
\end{equation}
and
\begin{equation}
\begin{split}
n_0^{} &= \frac{2k_F^3}{3\pi^2}\,, \quad E_{F}^{*} = \sqrt{k_F^2 + M^{*2}}\,,\\
S_0 &= g_{\sigma}^{} \sigma_0^{}\,, \quad W_0 = g_{\omega}^{} \omega_0^{}\,.
\end{split}
\end{equation}
Here, the subscript 0 indicates that the quantity is computed at the saturation density.
Once we know the saturation properties, such as $n_0$, $M^{*}$, and $B$ 
$( = -\mu_n^{} + M= - \mu_p^{} + M)$, the values of $k_F$, $S_0$, and $W_0$ can be
obtained from the above relations. 
In addition, the field equations for the $\sigma$ and the $\omega$ mesons are
\begin{align}
& \left(\frac{m_\sigma}{g_\sigma}\right)^2 S_0 
+ \frac{\kappa}{2}S_0^2 + \frac{\lambda}{6} S_0^{3}
- \frac{2}{\pi^2}\int_{0}^{k_F} dk \frac{k^2 M^*}{\sqrt{k^2 + M^{*2}}} =0 \,,
\label{eq:sigma}
\\
& \left(\frac{m_\omega}{g_\omega}\right)^2 W_0 
+ \frac{\zeta}{6}W_0^3 
- \frac{2}{3\pi^2}k_F^3 =0 \label{eq:omega}\,.
\end{align}
The five equations, \eqref{eq:bind}, \eqref{eq:pres}, \eqref{eq:comp}, \eqref{eq:sigma}, 
and \eqref{eq:omega}, are used to determine the values of five unknowns, 
$g_\sigma^{}$, $g_\omega^{}$, $\kappa$, $\lambda$, and $\zeta$. 
These equations, however, are redundant because of the relation on the pressure, 
$P = \mu_n^{} n_n + \mu_p^{} n_p - \mathcal{E}$. 
Therefore, we need one more information to determine the model parameters.
As in Sec.~\ref{sec:nuc}, however, we set $\zeta=0$ for allowing the simple rescaling of $g_{\phi N}$.%
\footnote{As shown in Table~\ref{tb:rmfpa}, the IU-FSU and SFHo models use $\zeta = 3.0 \times 10^{-2}$ and
$-1.701 \times 10^{-3}$, respectively.}

For pure neutron matter, we have eleven unknowns ($g_\rho^{}$, $\xi$, $\Lambda_{s1}$, $\dots$, $\Lambda_{s6}$,
$\Lambda_{v1}$, $\dots$, $\Lambda_{v3}$,) to be determined. 
As in IU-FSU~\cite{FHPS10}, we may use two coupling constants, i.e., $\Lambda_{s2} $ and $\Lambda_{v1}$, 
for neutron matter. 
It does not, however, give a good fit to pure neutron matter calculation. 
Therefore, we maintain the number of coupling constant as in Ref.~\cite{SPLE04}.

The explicit expressions for symmetry energy and its density derivative are obtained as
\begin{eqnarray}
S_v & = &
\frac{1}{8}\frac{\partial^2 (\mathcal{E}/n)}{\partial \alpha^2}\biggl\vert_{\alpha=0} 
\nonumber \\ 
& = & \frac{k_F^2}{6E_F^*} + \frac{n}{8 \left[ (m_\rho^2 / g_\rho^2 ) + 2f(S,W)\right]} \,,
\\
L & = & \frac{3n_0}{8}\frac{\partial^3 (\mathcal{E}/n)}{\partial n \partial \alpha^2}\biggl\vert_{\alpha=0} 
\nonumber \\
 & = & \frac{k_F^2}{6E_F^*}\left[
    1 + \frac{M^{*2}}{E_F^2} - \frac{3n_0 M^*}{E_F^2}\frac{\partial M^*}{\partial n} \right]  
\nonumber \\ &&  \mbox{}
+ \frac{3n_0}{8\left[ (m_\rho^2 / g_\rho^2 ) + 2f\right]} 
\nonumber \\  \nonumber \\ && \mbox{}
- \frac{3n_0^2}{4\left[ (m_\rho^2/ g_\rho^2 ) + 2f\right]^{2}}
    \left(\frac{\partial f}{\partial S}\frac{\partial S}{\partial n}
      + \frac{\partial f}{\partial W}\frac{\partial W}{\partial n}\right)\,,
\nonumber \\
\end{eqnarray}
where $\partial S/\partial n$ and $\partial W/\partial n$ are 
\begin{align}
\frac{\partial S}{\partial n} & = - \frac{\partial M^*}{\partial n} \\
& = \frac{M^*}{\sqrt{k_F^2 + M^{*2}}}
\left[
\frac{d^2 V}{d S^2}
+ \frac{2}{\pi^2}\int_0^{k_F}\frac{k^4}{(k^2 + M^{*2})^{3/2}}\,dk
\right]^{-1} \notag
\\
\frac{\partial W}{\partial n} & 
= \left[\frac{m_\omega^2}{g_\omega^2} + \frac{\zeta }{2}W_0^2\right]^{-1}.
\end{align}
In this derivation we utilize Eqs.~\eqref{eq:sigma} and \eqref{eq:omega}.


\begin{thebibliography}{10}

\bibitem{LP04b}
J.~M. Lattimer and M.~Prakash,
\newblock The physics of neutron stars,
\newblock Science \textbf{304}, 536 (2004).

\bibitem{Lattimer13}
J.~M. Lattimer,
\newblock The nuclear equation of state and neutron star masses,
\newblock Ann. Rev. Nucl. Part. Sci. \textbf{62}, 485 (2012).

\bibitem{DPRRH10}
P.~B. Demorest, T.~Pennucci, S.~M. Ransom, M.~S.~E. Roberts, and J.~W.~T.
  Hessels,
\newblock A two-solar-mass neutron star measured using Shapiro delay,
\newblock Nature \textbf{467}, 1081 (2010).

\bibitem{AFWT13}
J.~Antoniadis \textit{et~al.\/},
\newblock A massive pulsar in a compact relativistic binary,
\newblock Science \textbf{340}, 1233232 (2013).

\bibitem{Weber04}
F.~Weber,
\newblock Strange quark matter and compact stars,
\newblock Prog. Part. Nucl. Phys. \textbf{54}, 193 (2005).

\bibitem{ABPR04}
M.~Alford, M.~Braby, M.~Paris, and S.~Reddy,
\newblock Hybrid stars that masquerade as neutron stars,
\newblock Astrophys. J. \textbf{629}, 969 (2005).

\bibitem{WSPHS11}
S.~Weissenborn, I.~Sagert, G.~Pagliara, M.~Hempel, and J.~Schaffner-Bielich,
\newblock Quark matter in massive compact stars,
\newblock Astrophys. J. Lett. \textbf{740}, L14 (2011).

\bibitem{Baym73}
G.~Baym,
\newblock Pion condensation in nuclear and neutron star matter,
\newblock Phys. Rev. Lett. \textbf{30}, 1340 (1973).

\bibitem{AB74}
C.-K. Au and G.~Baym,
\newblock Pion condensation in neutron star matter (II). Nuclear forces and
  stability,
\newblock Nucl. Phys. A \textbf{236}, 500 (1974).

\bibitem{TPL93}
V.~Thorsson, M.~Prakash, and J.~M. Lattimer,
\newblock Composition, structure and evolution of neutron stars with kaon
  condensates,
\newblock Nucl. Phys. A \textbf{572}, 693 (1994),
\newblock \textbf{574}, 851(E) (1994).

\bibitem{GS98c}
N.~K. Glendenning and J.~Schaffner-Bielich,
\newblock Kaon condensation and dynamical nucleons in neutron stars,
\newblock Phys. Rev. Lett. \textbf{81}, 4564 (1998).

\bibitem{LKHL13}
Y.~Lim, K.~Kwak, C.~H. Hyun, and C.-H. Lee,
\newblock Kaon condensation in neutron stars with Skyrme-Hartree-Fock models,
\newblock Phys. Rev. C \textbf{89}, 055804 (2014).

\bibitem{VLPPB10}
I.~Vida{\~n}a, D.~Logoteta, C.~Provid\^{e}ncia, A.~Polls, and I.~Bombaci,
\newblock Estimation of the effect of hyperonic three-body forces on the
  maximum mass of neutron stars,
\newblock EPL \textbf{94}, 11002 (2011).

\bibitem{YFYR13}
Y.~Yamamoto, T.~Furumoto, N.~Yasutake, and \mbox{Th}. A.~Rijken,
\newblock Multi-Pomeron repulsion and the neutron-star mass,
\newblock Phys. Rev. C \textbf{88}, 022801(R) (2013).

\bibitem{YFYR14}
Y.~Yamamoto, T.~Furumoto, N.~Yasutake, and \mbox{Th}. A.~Rijken,
\newblock Hyperon mixing and universal many-body repulsion in neutron stars,
\newblock Phys. Rev. C \textbf{90}, 045805 (2014).

\bibitem{LLGP14}
D.~Lonardoni, A.~Lovato, S.~Gandolfi, and F.~Pederiva,
\newblock Hyperon puzzle: Hints from Quantum Monte Carlo calculations,
\newblock Phys. Rev. Lett. \textbf{114}, 092301 (2015).

\bibitem{BD00}
I.~Bombaci and B.~Datta,
\newblock Conversion of neutron stars to strange stars as the central engine of
  gamma-ray bursts,
\newblock Astrophys. J. \textbf{530}, L69 (2000).

\bibitem{BBDFL02}
Z.~Berezhiani, I.~Bombaci, A.~Drago, F.~Frontera, and A.~Lavagno,
\newblock Gamma-ray bursts from delayed collapse of neutron stars to quark
  matter stars,
\newblock Astrophys. J. \textbf{586}, 1250 (2003).

\bibitem{BPV04}
I.~Bombaci, I.~Parenti, and I.~Vida{\~{n}}a,
\newblock Quark deconfinement and implications for the radius and the limiting
  mass of compact stars,
\newblock Phys. Rev. D \textbf{614}, 314 (2004).

\bibitem{AS60}
V.~A. Ambartsumyan and G.~S. Saakyan,
\newblock The degenerate superdense gas of elementary particles,
\newblock Astron. Zh. \textbf{37}, 193 (1960),
\newblock [Soviet Astron. AJ \textbf{4}, 187--201 (1960)].

\bibitem{Pandharipande71}
V.~R. Pandharipande,
\newblock Hyperonic matter,
\newblock Nucl. Phys. A \textbf{178}, 123 (1971).

\bibitem{BJ74}
H.~A. Bethe and M.~B. Johnson,
\newblock Dense baryon matter calculations with realistic potentials,
\newblock Nucl. Phys. A \textbf{230}, 1 (1974).

\bibitem{Glendenning82}
N.~K. Glendenning,
\newblock The hyperon composition of neutron stars,
\newblock Phys. Lett. \textbf{114B}, 392 (1982).

\bibitem{GM91}
N.~K. Glendenning and S.~A. Moszkowski,
\newblock Reconciliation of neutron-star masses and binding of the $\Lambda$ in
  hypernuclei,
\newblock Phys. Rev. Lett. \textbf{67}, 2414 (1991).

\bibitem{BG04}
J.~K. Bunta and \v{S}tefan Gmuca,
\newblock Hyperons in a relativistic mean-field approach to asymmetric nuclear
  matter,
\newblock Phys. Rev. C \textbf{70}, 054309 (2004).

\bibitem{BM09b}
I.~Bednarek and R.~Manka,
\newblock The role of nonlinear vector meson interactions in hyperon stars,
\newblock J. Phys. G \textbf{36}, 095201 (2009).

\bibitem{BHZBM11}
I.~Bednarek, P.~Haensel, J.~L. Zdunik, M.~Bejger, and R.~Ma{\'n}ka,
\newblock Hyperons in neutron-star cores and a $2 M_\odot$ pulsar,
\newblock Astron. Astrophys. \textbf{543}, A157 (2012).

\bibitem{WCS12a}
S.~Weissenborn, D.~Chatterjee, and J.~Schaffner-Bielich,
\newblock Hyperons and massive neutron stars: The role of hyperon potentials,
\newblock Nucl. Phys. A \textbf{881}, 62 (2012).

\bibitem{WCS12b}
S.~Weissenborn, D.~Chatterjee, and J.~Schaffner-Bielich,
\newblock Hyperons and massive neutron stars: Vector repulsion and SU(3)
  symmetry,
\newblock Phys. Rev. C \textbf{85}, 065802 (2012),
\newblock \textbf{90}, 019904(E) (2014).

\bibitem{GHK14}
M.~E. Gusakov, P.~Haensel, and E.~M. Kantor,
\newblock Physics input for modelling superfluid neutron stars with hyperon
  cores,
\newblock Mon. Not. Roy. Astron. Soc. \textbf{439}, 318 (2014).

\bibitem{CV15}
D.~Chatterjee and I.~Vida{\~{n}}a,
\newblock Do hyperons exist in the interior of neutron stars?,
\newblock Eur. Phys. J. A \textbf{52}, 29 (2016).

\bibitem{BG97}
S.~Balberg and A.~Gal,
\newblock An effective equation of state for dense matter with strangeness,
\newblock Nucl. Phys. A \textbf{625}, 435 (1997).

\bibitem{LY97}
D.~E. Lanskoy and Y.~Yamamoto,
\newblock Skyrme-Hartree-Fock treatment of $\Lambda$ and $\Lambda\Lambda$
  hypernuclei with $G$-matrix motivated interactions,
\newblock Phys. Rev. C \textbf{55}, 2330 (1997).

\bibitem{GDS12}
N.~Guleria, S.~K. Dhiman, and R.~Shyam,
\newblock A study of $\Lambda$ hypernuclei within the Skyrme-Hartree-Fock
  model,
\newblock Nucl. Phys. A \textbf{886}, 71 (2012).

\bibitem{Mornas04}
L.~Mornas,
\newblock Neutron stars in a Skyrme model with hyperons,
\newblock Eur. Phys. J. A \textbf{24}, 293 (2005).

\bibitem{LHKL14}
Y.~Lim, C.~H. Hyun, K.~Kwak, and C.-H. Lee,
\newblock Hyperon puzzle of neutron stars with Skyrme force models,
\newblock Int. J. Mod. Phys. E \textbf{24}, 1550100 (2015).

\bibitem{OGP08}
F.~{\"O}zel, T.~G{\"u}ver, and D.~Psaltis,
\newblock The mass and radius of the neutron star in EXO 1745-248,
\newblock Astrophys. J. \textbf{693}, 1775 (2009).

\bibitem{SLB10}
A.~W. Steiner, J.~M. Lattimer, and E.~F. Brown,
\newblock The equation of state from observed masses and radii of neutron
  stars,
\newblock Astrophys. J. \textbf{722}, 33 (2010).

\bibitem{SPRW11}
V.~Suleimanov, J.~Poutanen, M.~Revnivtsev, and K.~Werner,
\newblock A neutron star stiff equation of state derived from cooling phases of
  the $X$-ray burster 4U 1724-307,
\newblock Astrophys. J. \textbf{742}, 122 (2011).

\bibitem{GSWR13}
S.~Guillot, M.~Servillat, N.~A. Webb, and R.~E. Rutledge,
\newblock Measurement of the radius of neutron stars with high signal-to-noise
  quiescent low-mass $X$-ray binaries in globular clusters,
\newblock Astrophys. J. \textbf{772}, 7 (2013).

\bibitem{OPGBHS16}
F.~\"{O}zel, D.~Psaltis, T.~G{\"u}ver, G.~Baym, C.~Heinke, and S.~Guillot,
\newblock The dense matter equation of state from neutron star radius and mass
  measurements,
\newblock Astrophys. J. \textbf{820}, 28 (2016).

\bibitem{LP06}
J.~M. Lattimer and M.~Prakash,
\newblock Neutron star observations: Prognosis for equation of state
  constraints,
\newblock Phys. Rep. \textbf{442}, 109 (2007).

\bibitem{LL12}
J.~M. Lattimer and Y.~Lim,
\newblock Constraining the symmetry parameters of the nuclear interaction,
\newblock Astrophys. J. \textbf{771}, 51 (2013).

\bibitem{LHL15}
Y.~Lim, C.~H. Hyun, and C.-H. Lee,
\newblock Nuclear equation of state and neutron star cooling,
\newblock Int. J. Mod. Phys. E \textbf{26}, 1750015 (2017).

\bibitem{OPAMB15}
F.~{\"O}zel, D.~Psaltis, Z.~Arzoumanian, S.~Morsink, and M.~Baub{\"o}ck,
\newblock Measuring neutron star radii via pulse profile modeling with NICER,
\newblock Astrophys. J. \textbf{832}, 92 (2016).

\bibitem{GCR11}
S.~Gandolfi, J.~Carlson, and S.~Reddy,
\newblock Maximum mass and radius of neutron stars, and the nuclear symmetry
  energy,
\newblock Phys. Rev. C \textbf{85}, 032801(R) (2012).

\bibitem{DSS13}
C.~Drischler, V.~Som{\`a}, and A.~Schwenk,
\newblock Microscopic calculations and energy expansions for neutron-rich
  matter,
\newblock Phys. Rev. C \textbf{89}, 025806 (2014).

\bibitem{HK16}
J.~W. Holt and N.~Kaiser,
\newblock Equation of state of nuclear and neutron matter at third-order in
  perturbation theory from chiral effective field theory,
\newblock Phys. Rev. C \textbf{95}, 034326 (2017).

\bibitem{RRH15}
E.~Rrapaj, A.~Roggero, and J.~W. Holt,
\newblock Microscopically constrained mean-field models from chiral nuclear
  thermodynamics,
\newblock Phys. Rev. C \textbf{93}, 065801 (2016).

\bibitem{SLB15}
A.~W. Steiner, J.~M. Lattimer, and E.~F. Brown,
\newblock Neutron star radii, universal relations, and the role of prior
  distributions,
\newblock Eur. Phys. J. A \textbf{52}, 18 (2016).

\bibitem{SPLE04}
A.~W. Steiner, M.~Prakash, J.~M. Lattimer, and P.~J. Ellis,
\newblock Isospin asymmetry in nuclei and neutron stars,
\newblock Phys. Rep. \textbf{411}, 325 (2005).

\bibitem{FHPS10}
F.~J. Fattoyev, C.~J. Horowitz, J.~Piekarewicz, and G.~Shen,
\newblock Relativistic effective interaction for nuclei, giant resonances, and
  neutron stars,
\newblock Phys. Rev. C \textbf{82}, 055803 (2010).

\bibitem{PDG16}
Particle Data Group, C.~Patrignani \textit{et~al.\/},
\newblock Review of particle physics,
\newblock Chin. Phys. C \textbf{40}, 100001 (2016).

\bibitem{GISPF09}
S.~Gandolfi, A.~\mbox{Yu}. Illarionov, K.~E. Schmidt, F.~Pederiva, and
  S.~Fantoni,
\newblock Quantum Monte Carlo calculation of the equation of state of neutron
  matter,
\newblock Phys. Rev. C \textbf{79}, 054005 (2009).

\bibitem{SHF12}
A.~W. Steiner, M.~Hempel, and T.~Fischer,
\newblock Core-collapse supernova equations of state based on neutron star
  observations,
\newblock Astrophys. J. \textbf{774}, 17 (2013).

\bibitem{LKR96}
G.~A. Lalazissis, J.~K{\"o}nig, and P.~Ring,
\newblock New parametrization for the Lagrangian density of relativistic mean
  field theory,
\newblock Phys. Rev. C \textbf{55}, 540 (1997).

\bibitem{SDGGS93}
J.~Schaffner, C.~B. Dover, A.~Gal, C.~Greiner, and H.~St{\"o}cker,
\newblock Strange hadronic matter,
\newblock Phys. Rev. Lett. \textbf{71}, 1328 (1993).

\bibitem{DG85}
C.~B. Dover and A.~Gal,
\newblock Hyperon-nucleus potentials,
\newblock Prog. Part. Nucl. Phys. \textbf{12}, 171 (1985).

\bibitem{SDGGS94}
J.~Schaffner, C.~B. Dover, A.~Gal, C.~Greiner, D.~J. Millener, and
  H.~St{\"o}cker,
\newblock Multiply strange nuclear systems,
\newblock Ann. Phys. (N.Y.) \textbf{235}, 35 (1994).

\bibitem{OK04}
Y.~Oh and H.~Kim,
\newblock Pentaquark baryons in the SU(3) quark model,
\newblock Phys. Rev. D \textbf{70}, 094022 (2004).

\bibitem{RSY99}
\mbox{Th}. A.~Rijken, V.~G.~J. Stoks, and Y.~Yamamoto,
\newblock Soft-core hyperon-nucleon potentials,
\newblock Phys. Rev. C \textbf{59}, 21 (1999).

\bibitem{ETR06}
G.~Erkol, R.~G.~E. Timmermans, and \mbox{Th}. A.~Rijken,
\newblock Vector-meson--baryon coupling constants in QCD sum rules,
\newblock Phys. Rev. C \textbf{74}, 045201 (2006).

\bibitem{BFSS99}
D.~Black, A.~H. Fariborz, F.~Sannino, and J.~Schechter,
\newblock Putative light scalar nonet,
\newblock Phys. Rev. D \textbf{59}, 074026 (1999).

\bibitem{SG00}
J.~Schaffner-Bielich and A.~Gal,
\newblock Properties of strange hadronic matter in bulk and in finite systems,
\newblock Phys. Rev. C \textbf{62}, 034311 (2000).

\bibitem{LH17}
Y.~Lim and J.~W. Holt,
\newblock Structure of neutron star crusts from new Skyrme effective
  interactions constrained by chiral effective field theory,
\newblock Phys. Rev. C \textbf{95}, 065805 (2017).

\bibitem{CBHMS98}
E.~Chabanat, P.~Bonche, P.~Haensel, J.~Meyer, and R.~Schaeffer,
\newblock A Skyrme parametrization from subnuclear to neutron star densities
  Part II. Nuclei far from stabilities,
\newblock Nucl. Phys. A \textbf{635}, 231 (1998),
\newblock \textbf{643}, 441(E) (1998).

\bibitem{HG94b}
P.~Haensel and O.~\mbox{Yu}. Gnedin,
\newblock Direct URCA processes involving hyperons and cooling of neutron
  stars,
\newblock Astron. Astrophys. \textbf{290}, 458 (1994).

\bibitem{SBS98}
C.~Schaab, S.~Balberg, and J.~Schaffner-Bielich,
\newblock Implications of hyperon pairing for cooling of neutron stars,
\newblock Astrophys. J. \textbf{504}, L99 (1998).

\bibitem{LHL16}
Y.~Lim, C.~H. Hyun, and C.-H. Lee,
\newblock Strangeness in neutron star cooling,
\newblock arXiv:1608.02078.

\bibitem{PPLP92}
M.~Prakash, M.~Prakash, J.~M. Lattimer, and C.~J. Pethick,
\newblock Rapid cooling of neutron stars by hyperons and Delta isobars,
\newblock Astrophys. J. \textbf{390}, L77 (1992).

\end{thebibliography}
\end{document}